\documentclass[times,twocolumn,tighten,trackchanges]{aastex62}

\shorttitle{Compound Chondrule Formation in Optically Thin Shock Waves}
\shortauthors{Arakawa \& Nakamoto}

\begin{document}

\title{Compound chondrule formation in optically thin shock waves
}

\correspondingauthor{Sota Arakawa}
\email{arakawa.s.ac@m.titech.ac.jp}

\author[0000-0003-0947-9962]{Sota Arakawa}
\affiliation{Department of Earth and Planetary Sciences, Tokyo Institute of Technology, Meguro, Tokyo 152-8551, Japan
}

\author[0000-0003-3924-6174]{Taishi Nakamoto}
\affiliation{Department of Earth and Planetary Sciences, Tokyo Institute of Technology, Meguro, Tokyo 152-8551, Japan
}

\begin{abstract}
Shock-wave heating within the solar nebula is one of the leading candidates for the source of chondrule-forming events.
Here, we examine the possibility of compound chondrule formation via optically thin shock waves.
Several features of compound chondrules indicate that compound chondrules are formed via the collisions of supercooled precursors.
We evaluate whether compound chondrules can be formed via the collision of supercooled chondrule precursors in the framework of the shock-wave heating model by using semi-analytical methods and discuss whether most of the crystallized chondrules can avoid destruction upon collision in the post-shock region.
We find that chondrule precursors immediately turn into supercooled droplets when the shock waves are optically thin and they can maintain supercooling until the condensation of evaporated fine dust grains.
Owing to the large viscosity of supercooled melts, supercooled chondrule precursors can survive high-speed collisions on the order of $1\ {\rm km}\ {\rm s}^{-1}$ when the temperature is below $\sim 1400\ {\rm K}$.
From the perspective of the survivability of crystallized chondrules, shock waves with a spatial scale of $\sim 10^{4}\ {\rm km}$ may be potent candidates for the chondrule formation mechanism.
Based on our results from one-dimensional calculations, a fraction of compound chondrules can be reproduced when the chondrule-to-gas mass ratio in the pre-shock region is $\sim 2 \times 10^{-3}$, which is approximately half of the solar metallicity.
\end{abstract}

\keywords{hydrodynamics --- shock waves --- meteorites, meteors, meteoroids --- protoplanetary disks}

\sloppy

\section{Introduction}

Chondrules are millimeter-sized spherical igneous grains contained within chondrites, which are the most common type of meteorites, as a major component.
The volume fraction of chondrules in ordinary chondrites is 60--80\% \citep[e.g.,][]{Rubin2000,Scott2007}, and the ages of chondrules are approximately 4.563--4.567 billion years, i.e., they were formed during the first 4 million years of the solar system \citep[e.g.,][]{Connelly+2012,Bollard+2017}.
Therefore, they must contain a wealth of information regarding the evolution of the solar nebula.
In the canonical view, small dust grains in the solar nebula grew into millimeter-sized aggregates, after which chondrules were formed by the melting of these aggregates in the early solar nebula and became spherical owing to their surface tension \citep[e.g.,][]{Zanda2004}; however, their precise origin is still unclear.

Some chondrules, referred to as compound chondrules, are composed of two or more chondrules fused together.
They comprise a low percentage of all chondrules \citep[e.g., 4\% in ordinary chondrites;][]{Gooding+1981}; however, they may offer crucial information regarding the physical state of solid materials during chondrule formation because they occur not only in ordinary chondrites but also in many classes of chondrites \citep[e.g.,][]{Akaki+2005,Bischoff+2017}.
Although the formation process of compound chondrules is still under debate, we can interpret the presence of compound chondrules as the result of collisions \citep[e.g.,][]{Gooding+1981,Ciesla+2004b,Miura+2008b,Bogdan+2019}.
The ubiquitous existence of cratered chondrules (approximately 10\% of all chondrules) also indicates that some of the chondrules have experienced collision when they crystallize \citep[e.g.,][]{Gooding+1981}.
\citet{Wasson+1995} examined compound chondrules in thin sections and classified each constituent chondrule as primary or secondary.
Primary chondrules retain their spherical shape, while secondary chondrules are deformed.
Compound chondrules with blurred intrachondrule boundaries are extremely rare within ordinary chondrites \citep{Wasson+1995}.
Therefore, most compound chondrules are formed by collisions between crystallized chondrules and non-crystallized precursors \citep{Arakawa+2016a}, or at least two components with a significant viscosity difference to be able to distinguish primary and secondary chondrules \citep{Yasuda+2009}.

Chondrules exhibit various textures, reflecting their different compositions and thermal histories \citep[e.g.,][]{Gooding+1981}.
In general, the textures of chondrules are classified into three textural types, that is, porphyritic, nonporphyritic, and glassy.
Porphyritic chondrules consist of phenocrysts of olivine and/or low-calcium pyroxene, with accessory amounts of sulfides and metal nuggets suspended in mesostasis.
Nonporphyritic chondrules are usually classified into three subtypes \citep[e.g.,][]{Gooding+1981}: cryptocrystalline, composed of nanometer- and micrometer-sized fine grains; radial-pyroxene; and barred-olivine chondrules (barred-pyroxene and radial-olivine chondrules also exist but are minor components).
Glassy chondrules are extremely rare, and they are only mentioned occasionally \citep[e.g.,][]{Krot+1994}.
It is typically thought that nonporphyritic and glassy chondrules are formed from completely molten precursors, while porphyritic chondrules melt incompletely during their formation \citep[e.g.,][]{Lofgren+1986,Hewins+1990}, although porphyritic textures can also be reproduced from completely molten precursors \citep[e.g.,][]{Connolly+1995,Srivastava+2010}.

Here, we note that the textures of chondrules contained in compound chondrules have noteworthy features.
\citet{Gooding+1981} and \citet{Wasson+1995} reported that approximately 15\% of all chondrules in ordinary chondrites are nonporphyritic, and most of them have porphyritic textures.
In contrast, when we observe the components in compound chondrules, most of the constituent chondrules are nonporphyritic \citep{Wasson1993,Wasson+1995}.
For the case of compound chondrules in ordinary chondrites, \citet{Wasson+1995} revealed that 81\% of primaries and 90\% of secondaries are nonporphyritic chondrules, and the same trend is also reported by \citet{Akaki+2005} for compound chondrules in CV carbonaceous chondrites.
Therefore, compound chondrules selectively form from precursors of nonporphyritic chondrules.
Dynamic crystallization experiments \citep[e.g.,][]{Tsukamoto+1999,Nagashima+2006,Nagashima+2008} have revealed that completely molten levitated precursors having no contact turn into supercooled droplets as they are cooled sufficiently below their liquidus temperature.
In addition, once these supercooled droplets collide with other particles, they crystallize instantaneously \citep[e.g.,][]{Connolly+1994}.
Therefore, when a crystallized chondrule and a supercooled precursor collide and stick together, a compound chondrule is formed \citep{Arakawa+2016a}.
This supercooled-collision scenario is consistent with the observed feature of the textures of chondrules contained in compound chondrules because the precursors of nonporphyritic chondrules selectively turn into supercooled droplets.

Numerous ideas have been proposed as mechanisms for single-chondrule formation, including shock-wave heating \citep[e.g.,][]{Hood+1991,Iida+2001,Boley+2013,Mai+2018}, planetesimal collisions \citep[e.g.,][]{Asphaug+2011,Dullemond+2014,Johnson+2015,Wakita+2017}, and radiative heating by lightning \citep[e.g.,][]{Horanyi+1995,Desch+2000,Muranushi2010,Johansen+2018}.
The combination of theoretical calculations and observations of chondrules provides several constraints on the properties of the chondrule formation mechanisms.
For example, the shapes of chondrules are usually close to perfect spheres, but some of them have prolate shapes \citep{Tsuchiyama+2003}; these prolate shapes can be explained by the rotation of molten chondrules exposed to a fast gas flow in the framework of the shock-wave heating model \citep{Miura+2008a}.
The maximum and minimum sizes of chondrules are also consistent with the theoretical predictions of shock-wave heating models \citep[e.g.,][]{Susa+2002,Miura+2005}.

Shock-wave heating within the solar nebula is one of the leading candidates for the source of chondrule-forming transient events.
Shock waves could be created by the eccentric planetesimals/protoplanets perturbed by Jovian resonances and the secular resonance caused by the gravity of the protoplanetary disk \citep[e.g.,][]{Weidenschilling+1998,Nagasawa+2019} or by gravitational instabilities in the protoplanetary disk \citep[e.g.,][]{Boss+2005,Boley+2008}.
The process of heating chondrule precursors by shock waves has been investigated in detail in many previous studies.
The shock-wave heating model can satisfy various first-order constraints related to chondrule formation, such as the peak temperature and the formation age \citep[e.g.,][]{Desch+2012}.

One important challenge for shock-wave heating models was noted by \citet{Nakamoto+2004} and \citet{Jacquet+2014}: chondrule precursors of different sizes have different velocities in the post-shock region, and they should collide at a high speed (approximately a few ${\rm km}\ {\rm s}^{-1}$), which may cause their destruction rather than compound chondrule formation upon collision.
However, the critical velocity for collisional sticking/destruction may strongly depend on the physical states of colliding precursors, e.g., phase, temperature, and size ratio.
For example, \citet{Ciesla2006} argued that partially molten chondrules with highly viscous outer layers could survive high-speed collisions because energy dissipation in droplet collisions increases as the viscosity of the liquid is increased \citep[e.g.,][]{Ennis+1991,Willis+2003}.
We note that the viscosity of silicate melts strongly depends on the temperature, and supercooled droplets must have significantly high viscosity \citep[e.g.,][]{Fulcher1925}; therefore, the collision of supercooled chondrule precursors in post-shock regions can potentially explain the formation of compound chondrules.
In addition, for the case of optically thin shock waves, collisions of chondrule precursors mostly occur when they are in the supercooled state.

In this study, we examine the possibility of compound chondrule formation via optically thin shock waves.
We evaluate whether compound chondrules can be formed via the collision of supercooled chondrule precursors in the framework of the shock-wave heating model by using semi-analytical methods and discuss whether most of the crystallized chondrules can avoid destruction upon collision in the post-shock region.
The objectives of this study are to postulate how the supercooling of chondrule precursors could affect the outcomes of high-speed collisions and suggest a novel scenario for compound chondrule formation.

\section{Models}

\subsection{Outline}
\label{sec2.1}

Most of the previous studies on chondrule-forming shock-wave heating models assumed that the shock waves are optically thick and chondrules are thermally coupled with gas in post-shock regions \citep[e.g.,][]{Morris+2010}; however, optically-thick shock waves have a critical issue in the context of compound chondrule formation.
Chondrules in optically thick shock waves should maintain a high temperature above their liquidus in post-shock regions \citep[e.g.,][]{Morris+2010}, and molten chondrules cannot avoid collisional destruction if they are in the molten state \citep[e.g.,][]{Jacquet+2014}.
Therefore, in this study, we examine the scenario whereby compound chondrules are formed via optically thin shock waves.
The prominent feature of the optically thin shock-wave model is its rapid cooling as a result of radiative cooling \citep[e.g.,][]{Ciesla+2004a}.

\begin{figure}
\centering
\includegraphics[width=0.9\columnwidth]{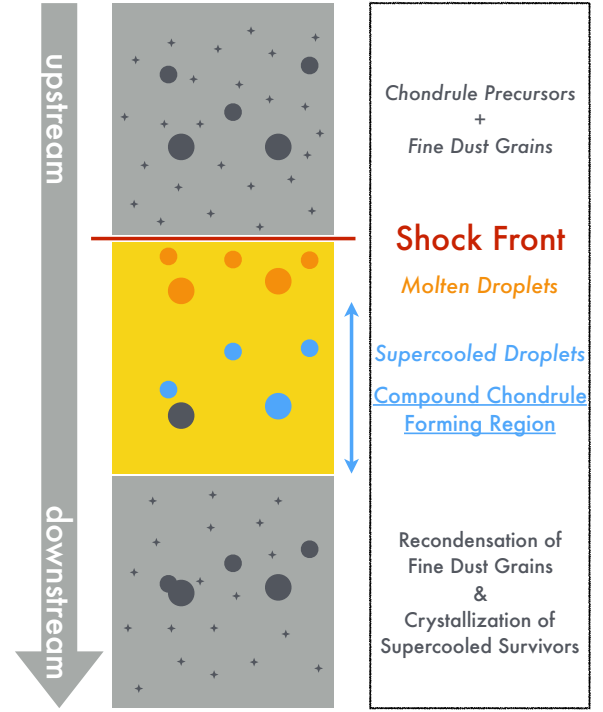}
\caption{
Outline of our compound chondrule formation scenario.
Molten chondrule precursors formed via passage of the shock front immediately turn into supercooled droplets because of their radiative cooling.
Then, some supercooled precursors experience collision and become crystallized chondrules, and if a crystallized chondrule and a supercooled precursor collide and stick together, a compound chondrule is formed.
}
\label{fig1}
\end{figure}

The formation process of single and compound chondrules in an optically thin shock wave is illustrated in Figure \ref{fig1}.
There are chondrule precursors and fine dust grains in the pre-shock region; the fine dust grains should evaporate immediately after passing the shock front, while the chondrule precursors are converted into molten droplets \citep[e.g.,][]{Miura+2005}.
There are no fine dust grains immediately behind the shock front, and these evaporated dust grains recondense when the gas temperature drops below the dust condensation temperature $T_{\rm c}$ (in this study, we assume $T_{\rm c} = 1600\ {\rm K}$).
Molten precursors formed via the passage of the shock front quickly transform into supercooled droplets because of their radiative cooling, and the temperature of supercooled droplets is controlled by the balance between the energy transfer from hot gas molecules to cold droplets and the radiative cooling of droplets (see Equation \ref{eqTeq}).
Although most of the precursors are in the supercooled state before the recondensation of fine dust grains, some precursors experience collision and become crystallized chondrules before the recondensation of fine dust grains.
Moreover, if a crystallized chondrule and a supercooled precursor collide and stick together, a compound chondrule is formed \citep{Arakawa+2016a}.
Finally, the gas temperature decreases and the recondensation of fine dust grains occurs downstream, after which supercooled survivors collide with fine dust grains and turn into crystallized chondrules.

\subsection{Chondrule dynamics}

In this study, we consider one-dimensional normal shocks, as in previous studies \citep[e.g.,][]{Nakamoto+2004,Ciesla2006,Jacquet+2014}.
We do not calculate the thermal/dynamical evolution of gas behind the shock front; we assume a simple gas structure, so that the dynamics of chondrules is simulated in the given gas flow.
We assume that the gas velocity with respect to the shock front $v_{\rm g}$ and the gas density $\rho_{\rm g}$ change across the shock front as functions of the distance from the shock front $x$ as follows:
\begin{equation}
v_{\rm g} = \left\{ \begin{array}{ll}
            v_{0} & {(x < 0)}, \\
            v_{0} + {\left( v_{\rm post} - v_{0} \right)} \exp{\left( {- x}/{L} \right)} & {(x \geq 0)},
            \end{array} \right.
\end{equation}
and
\begin{equation}
\rho_{\rm g} = \left\{ \begin{array}{ll}
            \rho_{{\rm g}, 0} & {(x < 0)}, \\
            {\left( {v_{\rm g}}/{v_{0}} \right)}^{-1} \rho_{{\rm g}, 0} & {(x \geq 0)},
            \end{array} \right.
\end{equation}
where $v_{0}$ is the pre-shock gas velocity with respect to the shock front, $v_{\rm post}$ is the post-shock gas velocity with respect to the shock front, $\rho_{{\rm g}, 0}$ is the pre-shock gas density, and $L$ is the spatial scale of the chondrule-forming shock.
The post-shock gas velocity, $v_{\rm post}$, is given by the Rankine--Hugoniot relations as $v_{\rm post} = {\left[ {(\gamma - 1)}/{(\gamma + 1)}\right]} v_{0}$, where $\gamma$ is the ratio of specific heats.
In this study, we set $\rho_{{\rm g}, 0} = 3 \times 10^{-9}\ {\rm g}\ {\rm cm}^{-3}$, $v_{0} = 12\ {\rm km}\ {\rm s}^{-1}$, and $\gamma = 1.4$.
Similarly, the temperature of the gas $T_{\rm g}$ is assumed as follows:
\begin{equation}
T_{\rm g} = \left\{ \begin{array}{ll}
            T_{0} & {(x < 0)}, \\
            T_{0} + {\left( T_{\rm post} - T_{0} \right)} \exp{\left( {- x}/{L} \right)} & {(x \geq 0)},
            \end{array} \right.
\label{eqTg}
\end{equation}
and we assume that the pre-shock gas temperature is $T_{0} = 500\ {\rm K}$ and the post-shock gas temperature is $T_{\rm post} = 2000\ {\rm K}$.
The sound velocity $c_{\rm s}$ is given by $c_{\rm s} \equiv {(2 k_{\rm B} T_{\rm g} / m_{\rm g})}^{1/2}$, where $k_{\rm B} = 1.38 \times 10^{-16}\ {\rm erg}\ {\rm K}^{-1}$ is the Boltzmann constant, and we set the gas molecule mass $m_{\rm g} = 3.34 \times 10^{-24}\ {\rm g}$, which values correspond to ${\rm H}_{2}$ gas.

The velocity of chondrules with respect to the shock front, $v$, will change as a function of the distance from the shock front $x$ \citep[e.g.,][]{Hood+1991}:
\begin{equation}
\frac{4 \pi}{3} r^{3} \rho \frac{{\rm d}v}{{\rm d}x} = - \frac{C_{\rm D}}{2} \pi r^{2} \rho_{\rm g} \frac{\left| v - v_{\rm g} \right|}{v} {\left( v - v_{\rm g} \right)},
\end{equation}
where $C_{\rm D}$ is the drag coefficient, $r$ is the chondrule radius, and $\rho = 3.3\ {\rm g}\ {\rm cm}^{-3}$ is the internal density of chondrules \citep{Ciesla+2004a}.
The drag coefficient $C_{\rm D}$ is given by
\begin{eqnarray}
C_{\rm D} &=& \frac{2}{3 s} \sqrt{\frac{\pi T}{T_{\rm g}}} + \frac{2 s^{2} + 1}{\sqrt{\pi} s^{3}} \exp{(- s^{2})} \nonumber \\
          && + \frac{4 s^{4} + 4 s^{2} - 1}{2 s^{4}} {\rm erf}{(s)},
\end{eqnarray}
where the temperature of the chondrule is $T$, and $s$ is given by $s \equiv {|v - v_{\rm g}|} / c_{\rm s}$.
The drag coefficient $C_{\rm D}$ is a function of the normalized relative velocity $s$, and we note that $C_{\rm D}$ approaches
\begin{equation}
C_{\rm D} \simeq 2,
\end{equation}
for the supersonic limit (i.e., $s \gg 1$) and
\begin{equation}
C_{\rm D} \simeq \frac{16 \sqrt{\pi}}{3 s} {\left( \sqrt{\frac{T}{T_{\rm g}}} + \frac{1}{8 \pi} \right)},
\end{equation}
for the subsonic limit (i.e., $s \ll 1$).
We can understand the dynamics of chondrules by considering the stopping length $l_{\rm stop}$.
For the case in which chondrules move in gas with supersonic velocities, $l_{\rm stop}$ is approximately given by
\begin{eqnarray}
l_{\rm stop} &\equiv& {\left( \frac{1}{v} \frac{{\rm d}v}{{\rm d}x} \right)}^{-1} \simeq \frac{4}{3} \frac{\rho}{\rho_{\rm g}} {\left( \frac{v - v_{\rm g}}{v} \right)}^{-2} r \nonumber \\
             &\sim& 2 \times 10^{2}\ {\left( \frac{v}{v - v_{\rm g}} \right)}^{2} {\left( \frac{r}{1\ {\rm mm}} \right)} \nonumber \\
             && \cdot {\left( \frac{\rho_{\rm g}}{2 \times 10^{-8}\ {\rm g}\ {\rm cm}^{-3}} \right)}^{-1}\ {\rm km}.
\label{eqlstop}
\end{eqnarray}
If the spatial scale of shock $L$ is much larger than $l_{\rm stop}$, the velocity of a chondrule $v$ reaches $v_{\rm post}$ behind the shock front, while $v$ barely changes when $L \ll l_{\rm stop}$ (see Figure \ref{fig3}b).

The equation of energy for a chondrule in gas is given by \citep[e.g.,][]{Hood+1991}
\begin{equation}
\frac{4 \pi}{3} r^{3} \rho c_{\rm heat} \frac{{\rm d}T}{{\rm d}x} = \frac{4 \pi r^{2}}{v} {\left( \Gamma - \Lambda \right)},
\end{equation}
where $c_{\rm heat} = 1 \times 10^{7}\ {\rm erg}\ {\rm g}^{-1}\ {\rm K}^{-1}$ is the specific heat \citep{Ciesla+2004a}, $\Gamma$ is the heating rate via gas--chondrule energy transfer per unit area, and $\Lambda$ is the rate of radiative cooling per unit area.
In this study, the effects of latent heat and evaporation \citep[e.g.,][]{Miura+2002} are not considered for simplicity.
The heating rate via gas--chondrule energy transfer $\Gamma$ is
\begin{equation}
\Gamma = \rho_{\rm g} {\left| v - v_{\rm g} \right|} {\left( T_{\rm rec} - T \right)} C_{\rm H},
\end{equation}
where $T_{\rm rec}$ is the adiabatic recovery temperature and $C_{\rm H}$ is the heat transfer function, called the Stanton number.
The adiabatic recovery temperature $T_{\rm rec}$ and the Stanton number $C_{\rm H}$ are given by \citep[e.g.,][]{Gombosi+1986}
\begin{eqnarray}
T_{\rm rec} &=& \frac{T_{\rm g}}{\gamma + 1} {\displaystyle\biggl [} 2 \gamma + 2 {\left( \gamma - 1 \right)} s^{2} \nonumber \\
            && - \frac{\gamma - 1}{{(1/2)} + s^{2} + {({s}/{\sqrt{\pi}})} \exp{(- s^{2})} {\rm erf}^{-1}{(s)}} {\displaystyle\biggl ]}, \nonumber \\
\end{eqnarray}
and
\begin{equation}
C_{\rm H} = \frac{\gamma + 1}{\gamma - 1} \frac{k_{\rm B}}{8 m_{\rm g} s^{2}} {\left[ \frac{s}{\sqrt{\pi}} \exp{(- s^{2})} + {\left( \frac{1}{2} + s^{2} \right)} {\rm erf}{(s)} \right]}.
\end{equation}

We assume that the optical depth of the chondrule-forming region is not far larger than unity and chondrules are thermally decoupled from the gas.
Here, we check the validity of this assumption.
We define ${\cal R}_{\rm w}$ as the width of the warm region (i.e., the region with a gas temperature of $T_{\rm g} \gg T_{0}$), and the width of the heating region whose optical depth is unity, ${\cal R}_{{\rm w}, 1}$, can be estimated as follows:
\begin{equation}
{\cal R}_{{\rm w}, 1} = {\left( \kappa \rho_{\rm g} \right)}^{-1} \sim 10^{3}\ {\left( \frac{\rho_{\rm g}}{3 \times 10^{-9}\ {\rm g}\ {\rm cm}^{-3}} \right)}^{-1}\ {\rm km},
\end{equation}
where $\kappa \sim 3\ {\rm cm}^{2}\ {\rm g}^{-1}$ is the opacity of the solar-metallicity protoplanetary disk \citep[e.g.,][]{Pollack+1985}. 
Therefore, if the width of the heating region ${\cal R}_{\rm w}$ is not larger than ${\cal R}_{{\rm w}, 1}$, and we do not consider significant enrichment of fine dust grains in the solar nebula, we can apply the optically thin approximation for chondrule-forming shock waves.
The width of the heating region ${\cal R}_{\rm w}$ is roughly given by the planetary radius ${\cal R}_{\rm p}$ when the shock waves are caused by eccentric planetary bodies \citep[e.g.,][]{Boley+2013}.
Under the optically thin shock assumption, the rate of radiative cooling per unit area of a chondrule $\Lambda$ is given by
\begin{equation}
\Lambda = \epsilon \sigma_{\rm SB} T^{4} - \epsilon \sigma_{\rm SB} T_{0}^{4},
\end{equation}
where $\epsilon = 0.9$ is the Planck mean emission/absorption coefficient \citep{Ciesla+2004a} and $\sigma_{\rm SB} = 5.67 \times 10^{-5}\ {\rm erg}\ {\rm cm}^{-2}\ {\rm K}^{-4}\ {\rm s}^{-1}$ is the Stefan-Boltzmann constant.

\subsection{Size-frequency distribution}

Several studies \citep[e.g.,][]{Rubin+1987,Nelson+2002,Metzler2018} have focused on chondrule size-frequency distributions.
The size-frequency distributions of chondrules usually use $\varnothing$-units, which are defined by,
\begin{equation}
\varnothing \equiv - \log_{2} \frac{2 r}{1\ {\rm mm}},
\end{equation}
or we can rewrite the above equation as $r = 2^{- {(\varnothing + 1)}}\ {\rm mm}$.
The mass of chondrules $m {(\varnothing)}$ is given by $m {(\varnothing)} = {\left( 4 \pi / 3 \right)} \rho r^{3}$.

Here, we assume that the size-frequency distribution in the pre-shock region ${f_{0} (\varnothing)}$ is similar to the size-frequency distribution in chondrites \citep{jacquet2014}; although \citet{Kadono+2005} proposed that the size-frequency distribution may originate from the breakup of huge molten precursors.
The size-frequency distribution of chondrules in ordinary chondrites is approximately log-normal \citep[e.g.,][]{Rubin+1987,Nelson+2002},
\begin{equation}
{f_{0} (\varnothing)} \propto \exp{\left[ - \frac{1}{2} {\left( \frac{\varnothing - \varnothing_{\rm mean}}{\varnothing_{\rm SD}} \right)}^{2} \right]};
\end{equation}
although, in reality, it is known that there is a cutoff for small chondrule sizes \citep[e.g.,][]{Eisenhour1996,Metzler2018}.
In this study, we assume $\varnothing_{\rm mean} = 0.8$ and $\varnothing_{\rm SD} = 0.8$, which are the mean and deviation for chondrules in LL ordinary chondrites \citep{Nelson+2002}.
The total number density of chondrules in pre-shock region $N_{0}$ is given by
\begin{equation}
N_{0} = \frac{\rho_{{\rm c}, 0}}{\int_{\varnothing_{\rm min}}^{\varnothing_{\rm max}} {\rm d}\varnothing\ {f_{0} (\varnothing)} {m (\varnothing)}},
\end{equation}
where $\rho_{{\rm c}, 0}$ is the mass density of chondrules in the pre-shock region, and $\varnothing_{\rm min}$ and $\varnothing_{\rm max}$ are the minimum and maximum of $\varnothing$ in the size-frequency distribution, respectively (in this study, we set $\varnothing_{\rm min} = -3$ and $\varnothing_{\rm max} = +3$).
The size-frequency distribution in the pre-shock region satisfies $\int_{\varnothing_{\rm min}}^{\varnothing_{\rm max}} {\rm d}\varnothing\ f_{0} = 1$ by definition.
The number density of chondrules whose size is $\varnothing$, ${n_{0} (\varnothing)}$, is also given by
\begin{equation}
{n_{0} (\varnothing)} = {f_{0} (\varnothing)} N_{0}.
\end{equation}
The number density of chondrules in the post-shock region, ${n (\varnothing, x)}$, changes with changing chondrule velocity $v = v {(\varnothing, x)}$.
Under the one-dimensional normal shock approximation, ${n (\varnothing, x)}$ is given as follows:
\begin{equation}
{n (\varnothing, x)} = {n_{0} (\varnothing)} \frac{v_{0}}{v (\varnothing, x)}.
\end{equation}

Using the geometrical optics approximation, the mean opacity of chondrules, $\kappa_{\rm c}$, is given by
\begin{equation}
\kappa_{\rm c} = \frac{\int_{\varnothing_{\rm min}}^{\varnothing_{\rm max}} {\rm d}\varnothing\ {n (\varnothing, x)} \pi r^{2} }{\int_{\varnothing_{\rm min}}^{\varnothing_{\rm max}} {\rm d}\varnothing\ {n (\varnothing, x)} m},
\end{equation}
and $\kappa_{\rm c}$ in the pre-shock region is $\kappa_{{\rm c}, 0} = 3.67\ {\rm cm}^{2}\ {\rm g}^{-1}$.
We found that $\kappa_{\rm c}$ is dominated by 0.5 mm-sized chondrules in the pre-shock region.
When we take into account the contribution of $\kappa_{\rm c}$, the optical depth of the heating region, $\tau$, is evaluated from
\begin{equation}
\tau = {\left( \kappa \rho_{\rm g} + \kappa_{\rm c} \rho_{\rm c} \right)} {\cal R}_{\rm w},
\end{equation}
the latter term, $\kappa_{\rm c} \rho_{\rm c}$, is negligibly smaller than the former term, $\kappa \rho_{\rm g}$, however.

\subsection{Collision frequency}

Here, we describe how to calculate the collision frequency of chondrules.
We define $\zeta_{\rm t,p}$ as the collision frequency per unit distance of a target chondrule, whose size is $\varnothing_{\rm t}$, with a projectile chondrule, whose size is $\varnothing_{\rm p}$.
Then, $\zeta_{\rm t,p}$ is given as follows:
\begin{eqnarray}
{\zeta_{\rm t,p} (\varnothing_{\rm t}, \varnothing_{\rm p}, x)} &=& {n (\varnothing_{\rm p}, x)} \cdot \pi {( r_{\rm t} + r_{\rm p} )}^{2} \nonumber \\
                                                                && \cdot \frac{|{v (\varnothing_{\rm p}, x) - v (\varnothing_{\rm t}, x)}|}{v (\varnothing_{\rm t}, x)}.
\end{eqnarray}
The collision frequency of a target chondrule with any projectile, $Z_{\rm t}$, is therefore given by
\begin{equation}
{Z_{\rm t} (\varnothing_{\rm t}, x)} = \int_{\varnothing_{\rm min}}^{\varnothing_{\rm max}} {\rm d}\varnothing_{\rm p}\ {\zeta_{\rm t,p} (\varnothing_{\rm t}, \varnothing_{\rm p}, x)}.
\end{equation}
Finally, the expected number of collisions for each target chondrule after passing the shock front, $\Sigma_{\rm t}$, is given by
\begin{equation}
{\Sigma_{\rm t} (\varnothing_{\rm t}, x)} = \int_{0}^{x} {\rm d}x'\ {Z_{\rm t} (\varnothing_{\rm t}, x')}.
\end{equation}

Here, we note that the fraction of compound chondrules among all the nonporphyritic chondrules in ordinary chondrites is approximately 20\% \citep{Ciesla+2004b,Arakawa+2016a}.
Therefore, the expected number of collisions $\Sigma_{\rm t}$ should be on the order of 20\% for small chondrules whose radii are comparable to that of typical secondaries, and $\Sigma_{\rm t}$ may be $\sim 1$--$2$ for large chondrules whose radii are comparable to that of typical primaries because primaries have experienced collisions twice \citep[see Section \ref{sec2.1} and][]{Arakawa+2016a}.

\subsection{Critical velocity for collisional sticking/merging}

When a droplet collides with a solid sphere, the expected collision outcomes are sticking, bouncing, or splashing \citep[e.g.,][]{Josserand+2016}.
Similarly, when two droplets collide, the collision outcomes are merging, bouncing/separation, or splashing \citep[e.g.,][]{Qian+1997}.
Bouncing usually occurs for grazing collisions.
In this study, we examine the critical velocity for compound chondrule formation from the view point of whether supercooled droplets can stick or not.
For the description of droplet collisions, it is necessary to consider the physical properties involved: viscosity $\eta$, density $\rho$, and surface tension $\sigma$, as well as geometrical properties of the system such as droplet radius $r$ and the impact velocity $v_{\rm imp}$.

\subsubsection{Dimensionless parameters for describing droplet collisions}

Using dimensional analysis, we can easily identify the relevant dimensionless parameters to describe binary collisions of liquid droplets \citep[e.g.,][]{Ashgriz+1990}.
For the case of head-on collision of equal-sized droplets with identical liquids, the basic parameters are the Weber number ${\rm We}$, the Reynolds number ${\rm Re}$, and the capillary number ${\rm Ca}$:
\begin{equation}
{\rm We} \equiv \frac{2 \rho r {v_{\rm imp}}^{2}}{\sigma},
\end{equation}
\begin{equation}
{\rm Re} \equiv \frac{2 \rho r v_{\rm imp}}{\eta},
\end{equation}
\begin{equation}
{\rm Ca} \equiv \frac{\eta v_{\rm imp}}{\sigma} \equiv \frac{{\rm We}}{{\rm Re}}.
\end{equation}
For the inviscid fluid limit (i.e., ${\rm Ca} \ll 1$), the criteria for collisional sticking should be given by the critical value of the Weber number ${\rm We}_{\rm cr, i}$;
\begin{equation}
{\rm We} < {\rm We}_{\rm cr, i},
\end{equation}
and, for the viscous fluid limit (i.e., ${\rm Ca} \gg 1$), the criteria should be given by the critical value of the Reynolds number ${\rm Re}_{\rm cr, v}$;
\begin{equation}
{\rm Re} < {\rm Re}_{\rm cr, v}.
\end{equation}
This expression can be converted into the expression of the critical Weber number by using the capillary number as follows:
\begin{equation}
{\rm We} < {\rm Re}_{\rm cr, v} {\rm Ca}.
\end{equation}
Therefore, we can imagine that the critical Weber number for collisional sticking, ${\rm We}_{\rm cr}$, can be given by the following equation:  
\begin{equation}
{\rm We}_{\rm cr} \simeq {\rm Re}_{\rm cr, v} {\rm Ca} + {\rm We}_{\rm cr, i}.
\label{eqDA}
\end{equation}

\subsubsection{Criteria proposed by \citet{Sommerfeld+2016}}
\label{sec2.5.2}

Droplets are affected by large deformation and energy dissipation when they collide; therefore, it is logical to use ${\rm Ca}$, which is the ratio of viscous forces to surface tension forces, for the expression of ${\rm We}_{\rm cr}$.
Recently, \citet{Sommerfeld+2016} proposed an equation for ${\rm We}_{\rm cr}$ as follows:
\begin{equation}
{\rm We}_{\rm cr} = \frac{K^{3}}{3} {\rm Ca} + 2 K,
\label{eqSK2016}
\end{equation}
where $K = 6.9451$ is called the structure parameter \citep{Naue+1992}, and we obtain ${\rm We}_{\rm cr} = 111.66 {\rm Ca} + 13.89$ (see Appendix \ref{ddc}).

From Equation (\ref{eqSK2016}), we can calculate the critical velocity for head-on collision of equal-sized droplets $v_{\rm cr}$ as follows:
\begin{equation}
v_{\rm cr} {(\eta, r)}= \frac{K^{3}}{12} \frac{\eta}{\rho r} {\left( 1 + \sqrt{1 + \frac{144}{K^{5}} \frac{\rho \sigma r}{{\eta}^{2}}} \right)},
\end{equation}
and when $v_{\rm cr}$ are controlled by viscous dissipation, these critical velocities are given by $v_{\rm cr} \sim 55.8 {\eta}/{(\rho r)}$.
In this case, the critical velocities are proportional to the viscosity and inversely proportional to the droplet radius.

For the case of collisions of different-sized droplets with different viscosities, the critical velocity for collisional merging $v_{\rm merge}$ is not yet understood \citep{Li+2016}.
In this study, we evaluate $v_{\rm merge}$ from the geometric mean of $v_{\rm cr}$ of the target and projectile:
\begin{equation}
v_{\rm merge} = \sqrt{v_{\rm cr} {(\eta_{\rm t}, r_{\rm t})} \cdot v_{\rm cr} {(\eta_{\rm p}, r_{\rm p})}},
\label{eqVm}
\end{equation}
where $\eta_{\rm t}$ and $\eta_{\rm p}$ are the viscosities of the target and projectile, respectively.
Likewise, when a non-crystallized projectile collides with a solidified target chondrule, we evaluate the critical velocity for collisional sticking $v_{\rm stick}$ from $v_{\rm cr}$ of the projectile (see Appendix \ref{dsc}):
\begin{equation}
v_{\rm stick} = v_{\rm cr} {(\eta_{\rm p}, r_{\rm p})}.
\label{eqVcr}
\end{equation}
The colliding supercooled droplets can turn into compound chondrules when the impact velocity $v_{\rm imp}$ is lower than $v_{\rm stick}$.
We note that our evaluation of $v_{\rm merge}$ and $v_{\rm stick}$ is not more than a rough order-of-magnitude estimate, and future studies on this issue are needed.

To determine the critical velocity for collisional sticking/merging, we need to know the material properties of silicate melts, $\eta$ and $\sigma$.
\citet{Hubbard2015} calculated the viscosities of chondrule melts by using the formula of \citet{Giordano+2008} which is based on the Vogel--Fulcher--Tammann viscosity equation \citep[][]{Vogel1921,Fulcher1925,Tammann+1926};
\begin{equation}
\log_{10}{\frac{\eta}{1\ {\rm P}}} = -3.55 + \frac{5084.9\ {\rm K}}{T - 584.9\ {\rm K}}.
\label{eqH2015}
\end{equation}
In contrast, the surface energy is only slightly dependent on the temperature, and we set $\sigma = 400\ {\rm erg}\ {\rm cm}^{-2}$ \citep{Murase+1973}.

\begin{figure}
\centering
\includegraphics[width=\columnwidth]{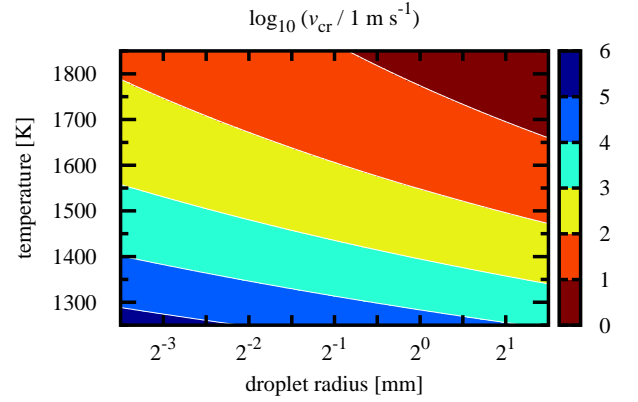}
\caption{
The critical velocity $v_{\rm cr}$ as a function of the radius of colliding droplets $r$ and their temperature $T$.
The viscosity of chondrule droplets is obtained from Equation (\ref{eqH2015}).
}
\label{fig2}
\end{figure}

The calculated $v_{\rm cr}$ is shown in Figure \ref{fig2}.
There is a strong dependence of $v_{\rm cr}$ on $T$, and we found that supercooled chondrule precursors could survive high-speed collisions on the order of $1\ {\rm km}\ {\rm s}^{-1}$ when the temperature is below $1400\ {\rm K}$.

\subsubsection{Temperature increase after collision}

When a droplet collides and sticks with another chondrule, the kinetic energy of the droplet is converted into thermal energy.
The impact energy $E_{\rm imp}$ and the thermal energy $E_{\rm th}$ are given by
\begin{equation}
E_{\rm imp} = \frac{1}{2} \frac{m_{\rm t} m_{\rm p}}{m_{\rm t} + m_{\rm p}} {v_{\rm imp}}^{2},
\end{equation}
and
\begin{equation}
E_{\rm th} = m_{\rm p} \rho c_{\rm heat} {\Delta T},
\end{equation}
where $m_{\rm t}$ and $m_{\rm p}$ are the masses of the target and the projectile, respectively.
By assuming $E_{\rm th} \simeq E_{\rm imp}$, the increase in the droplet temperature ${\Delta T}$ is estimated as follows:
\begin{equation}
{\Delta T} \sim 1.5 \times 10^{2}\ {\left( 1 + \frac{m_{\rm p}}{m_{\rm t}} \right)}^{-1} {\left( \frac{v_{\rm imp}}{1\ {\rm km}\ {\rm s}^{-1}} \right)}^{2}\ {\rm K}.
\end{equation}
This order estimation implies that, when the impact velocity $v_{\rm imp}$ is far larger than a few $ {\rm km}\ {\rm s}^{-1}$ and the projectile-to-target mass ratio $m_{\rm p} / m_{\rm t}$ is lower than unity, the colliding supercooled droplet would evaporate after collision rather than turn into a compound chondrule because the increase in the droplet temperature would be ${\Delta T} \gtrsim 1000\ {\rm K}$ (although we should consider the effect of the latent heat in reality).
Conversely, the effect of ${\Delta T}$ is negligible when $v_{\rm imp} \ll 1\ {\rm km}\ {\rm s}^{-1}$ or $m_{\rm p} / m_{\rm t} \gg 1$.
Hence we do not consider an increase in temperature after collision for simplicity.

\subsection{Catastrophic disruption criteria}

It is known that fragments of chondrules are common in chondrites \citep{Nelson+2002}, and fragmentation could have occurred in the solar nebula; for example, chondrule fragments within enveloping compound chondrules are fragmented in the solar nebula \citep{Wasson+1995}.
After the crystallization of chondrule precursors, the disruption of chondrules could occur in the post-shock region via high-speed collisions.
The catastrophic disruption criteria is $E_{\rm imp} \leq {\left( m_{\rm t} + m_{\rm p} \right)} Q_{\rm RD}^{*}$, and the critical specific energy for catastrophic disruption $Q_{\rm RD}^{*}$ is given by \citep{Stewart+2009}
\begin{equation}
Q_{\rm RD}^{*} = q_{\rm s} {\left( \frac{r_{\rm C1}}{1\ {\rm cm}} \right)}^{9 \mu / {(3 - 2 \varphi)}} {\left( \frac{v_{\rm imp}}{1\ {\rm cm}\ {\rm s}^{-1}} \right)}^{2 - 3 \mu}\ {\rm erg}\ {\rm g}^{-1},
\end{equation}
where $q_{\rm s}$, $\mu$, and $\varphi$ are dimensionless material properties, and the normalized radius $r_{\rm C1}$ is given as follows:
\begin{equation}
r_{\rm C1} = {\left( \frac{3}{4 \pi} \frac{m_{\rm t} + m_{\rm p}}{1\ {\rm g}\ {\rm cm}^{-3}} \right)}^{1/3}.
\end{equation}
For intact rocks such as basalt and granite, \citet{Stewart+2009} reported that the dimensionless material properties of $q_{\rm s} = 7 \times 10^{4}$, $\mu = 0.5$, and $\varphi = 8$ provide a reasonable fit for the experimental data.
Therefore the critical velocity for catastrophic disruption, $v_{\rm disrupt}$, is given as follows:
\begin{eqnarray}
v_{\rm disrupt} &=& 5.12 \times 10^{3}\ {\left( \frac{m_{\rm t}}{m_{\rm p}} \right)}^{2/3} {\left( 1 + \frac{m_{\rm p}}{m_{\rm t}} \right)}^{49/39} \nonumber \\
                && \cdot {\left( \frac{m_{\rm t}}{10^{-3}\ {\rm g}} \right)}^{- 1/13}\ {\rm cm}\ {\rm s}^{-1}.
\label{eqVdis}
\end{eqnarray}

Collisional disruption experiments with chondrules in Allende CV3 chondrite have been performed by \citet{Ueda+2001}, and they revealed that the catastrophic disruption criteria for similar-sized chondrules is approximately $1.5 \times 10^{4}\ {\rm cm}\ {\rm s}^{-1}$.
This experimental result validates our evaluation of $v_{\rm disrupt}$.

\section{Results}

\subsection{Chondrule dynamics and thermal history}
\label{sec3.1}

\begin{figure}
\centering
\includegraphics[width=\columnwidth]{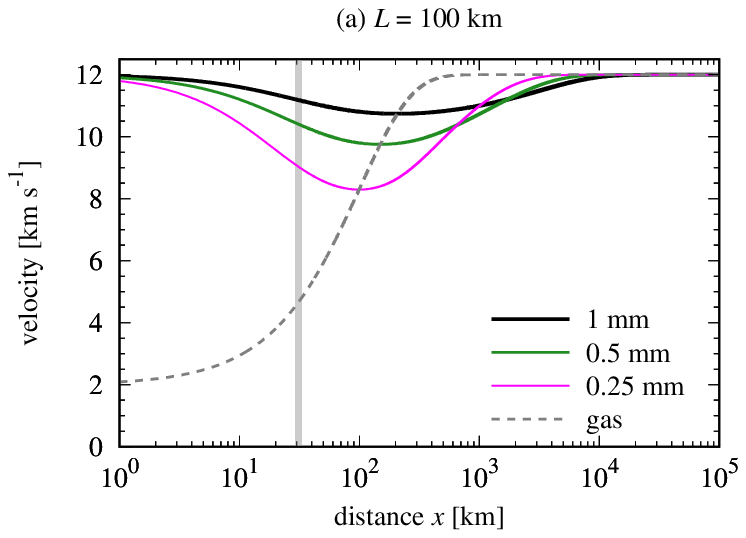}
\includegraphics[width=\columnwidth]{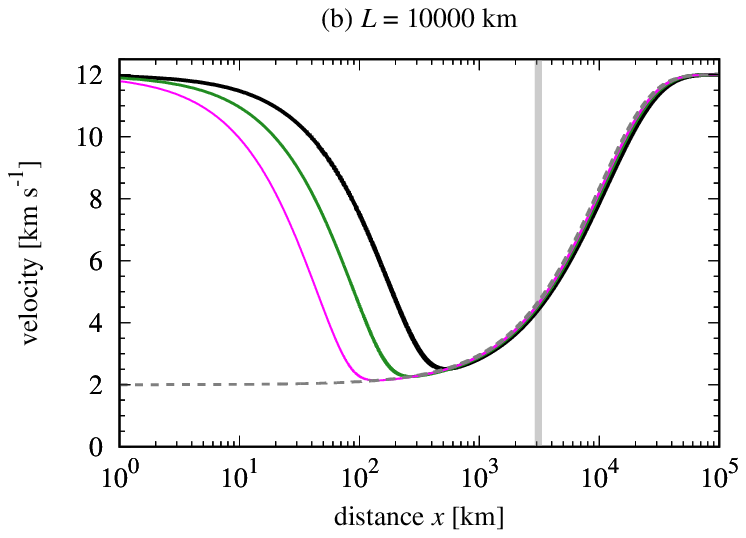}
\caption{
The velocity of chondrules with respect to the shock front $v$ and the gas velocity $v_{\rm g}$.
(a) For the case of the small-scale shock wave ($L = 100\ {\rm km}$).
(b) For the case of the large-scale shock wave ($L = 10000\ {\rm km}$).
The solid curves represent the velocity of chondrules with radii of $r = 1\ {\rm mm}$ (black), $r = 0.5\ {\rm mm}$ (green), and $r = 0.25\ {\rm mm}$ (magenta), and the gray dashed curve is the gas velocity.
The gray vertical line represents the recondensation line of evaporated fine dust grains.
}
\label{fig3}
\end{figure}

We first show the chondrule dynamics and thermal history in optically thin shock waves.
Here, we consider small and large shock waves whose spatial scales are $L = 100\ {\rm km}$ and $L = 10000\ {\rm km}$, respectively.
Figure \ref{fig3} shows the velocity of chondrules with respect to the shock front $v$ and gas velocity $v_{\rm g}$.
Figure \ref{fig3}a clearly shows that $v$ does not approach $v_{\rm post}$ ($= 2\ {\rm km}\ {\rm s}^{-1}$) for the small-scale shock wave.
In contrast, for the large-scale shock wave (Figure \ref{fig3}b), $v$ approaches $v_{\rm post}$ in the post-shock region.
This is because the stopping length of chondrules $l_{\rm stop}$ is significantly smaller than the spatial scale $L$ (see Equation \ref{eqlstop}).
For the case of Figure \ref{fig3}b, both $v$ and $v_{\rm g}$ change simultaneously when the distance from the shock front $x$ is larger than $1000\ {\rm km}$.
We derive an analytical equation of the chondrule-to-gas relative velocity, $v - v_{\rm g}$, in Section \ref{sec3.5}.

\begin{figure}
\centering
\includegraphics[width=\columnwidth]{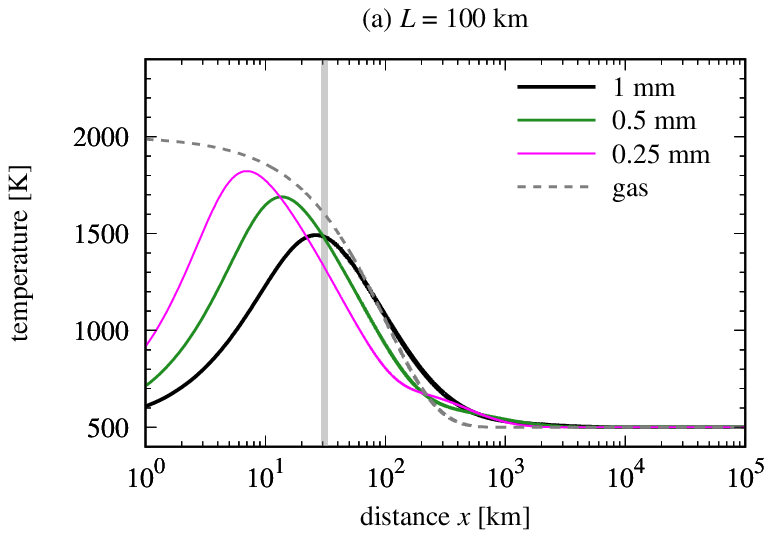}
\includegraphics[width=\columnwidth]{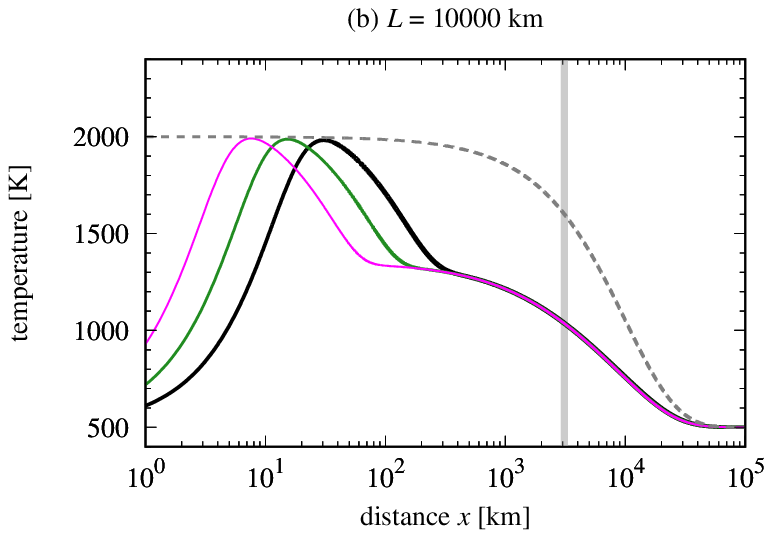}
\caption{
The temperature of chondrules $T$ and the gas temperature $T_{\rm g}$.
(a) For the case of the small-scale shock wave ($L = 100\ {\rm km}$).
(b) For the case of the large-scale shock wave ($L = 10000\ {\rm km}$).
The solid curves represent the temperature of chondrules with radii of $r = 1\ {\rm mm}$ (black), $r = 0.5\ {\rm mm}$ (green), and $r = 0.25\ {\rm mm}$ (magenta), and the gray dashed curve is the gas temperature.
The gray vertical line represents the recondensation line of evaporated fine dust grains.
}
\label{fig4}
\end{figure}

Figure \ref{fig4} is the temperature of chondrules $T$ and the gas temperature $T_{\rm g}$.
The gray vertical line represents the recondensation line of evaporated fine dust grains (i.e., the location where the gas temperature is $T_{\rm g} = T_{\rm c}$).
In this study, we set the condensation temperature to $T_{\rm c} = 1600\ {\rm K}$.

The liquidus temperature of chondrules is in the range of $1600$--$2100\ {\rm K}$ \citep[e.g.,][]{Cohen+2000}, and the solidus temperature is approximately $1400\ {\rm K}$ \citep[e.g.,][]{Sanders+2012}, although these temperatures depend on the composition of chondrules and the ambient pressure.
When the peak temperature of a chondrule is higher than the solidus temperature but lower than the liquidus temperature, the chondrule turns into a partially molten droplet.
In contrast, when the peak temperature is higher than the liquidus temperature, the chondrule becomes a completely molten droplet.
For the case of small-scale shock waves (Figure \ref{fig4}a), most of the small chondrules with radii of $r < 0.25\ {\rm mm}$ turn into completely molten precursors, while almost all large chondrules with radii of $r > 0.5\ {\rm mm}$ become partially molten droplets.

Several experimental studies have revealed that completely molten precursors turn into supercooled droplets and finally become glassy chondrules unless they collide with other particles \citep[e.g.,][]{Nagashima+2008}.
However, the observations of chondrules in a thin section revealed that glassy chondrules are extremely rare \citep[e.g.,][]{Krot+1994}.
This fact indicates that the recondensation of evaporated fine dust grains must occur before the temperature of supercooled precursors drops below the glass transition temperature $T_{\rm glass}$.
The glass transition temperature is dependent on the chemical composition, but it may be approximately $T_{\rm glass} \sim 900$--$1000\ {\rm K}$ \citep[e.g.,][]{Villeneuve+2015}.
Figure \ref{fig4}a shows that the recondensation of evaporated fine dust grains occurs before the temperature of supercooled precursors drops below the glass transition point; therefore, they can turn into crystallized chondrules without a glass transition.

The heating/cooling history of chondrules in the large-scale shock wave is shown in Figure \ref{fig4}b.
As in Figure \ref{fig4}a, recondensation of evaporated fine dust grains occurs before the temperature of supercooled precursors drops below the glass transition temperature, and these supercooled precursors can avoid turning into glassy chondrules.
The peak temperature of chondrules only slightly depends on their radii for the case of large-scale shock waves, and they can maintain the supercooling state for a long time.

\subsection{Equilibrium temperature of chondrules}

\begin{figure}
\centering
\includegraphics[width=\columnwidth]{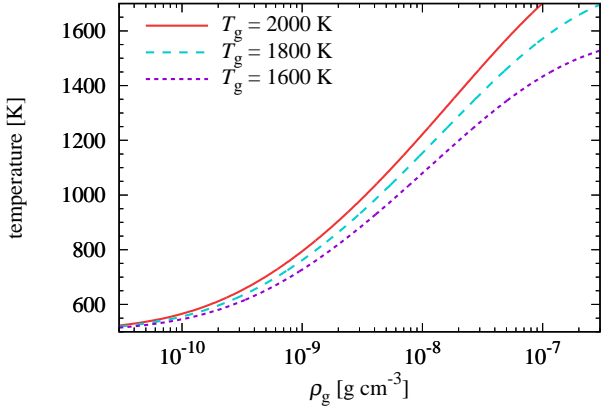}
\caption{
The temperature of chondrule precursors $T$ as a function of the gas density and the gas temperature, $\rho_{\rm g}$ and $T_{\rm g}$, under the assumption of $v - v_{\rm g} = 0$.
The temperature of chondrule precursors is calculated from Equation (\ref{eqTeq}) and we set $T_{0} = 500\ {\rm K}$.
}
\label{fig5}
\end{figure}

After the chondrule-to-gas relative velocity reaches zero (i.e., $s \to 0$), the temperature of chondrules in high-temperature gas can be calculated from the balance of the heating via collisions of high-temperature gas molecules and the radiative cooling of chondrules.
The heating term is given by
\begin{equation}
\Gamma = \frac{1}{8 \sqrt{\pi}} \frac{\gamma + 1}{\gamma - 1} \rho_{\rm g} c_{\rm s}^{3} {\left( 1 - \frac{T}{T_{\rm g}} \right)} \equiv \Gamma_{\rm g} {\left( 1 - \frac{T}{T_{\rm g}} \right)},
\end{equation}
and the cooling term is
\begin{equation}
\Lambda = \epsilon \sigma_{\rm SB} {\left( T^{4} - T_{0}^{4} \right)} = \Lambda_{\rm g} {\left[ {\left( \frac{T}{T_{\rm g}} \right)}^{4} - {\left( \frac{T_{0}}{T_{\rm g}} \right)}^{4} \right]},
\end{equation}
where $\Lambda_{\rm g} \equiv \epsilon \sigma_{\rm SB} T_{\rm g}^{4}$.
Then, we obtain the equilibrium value of $T$ by solving the equation, $\Gamma - \Lambda = 0$, and we can rewrite this equation as follows:
\begin{equation}
\frac{{(T/T_{\rm g})}^{4} - {(T_{0}/T_{\rm g})}^{4}}{1 - T/{T_{\rm g}}} = \frac{\Gamma_{\rm g}}{\Lambda_{\rm g}}.
\label{eqTeq}
\end{equation}
We find that there are two limiting cases; one case is that $T/T_{\rm g} \to 1$ and ${\Gamma_{\rm g}}/{\Lambda_{\rm g}} \to \infty$, and the other case is that $T/T_{\rm g} \to T_{0}/T_{\rm g}$ and ${\Gamma_{\rm g}}/{\Lambda_{\rm g}} \to 0$.
The dimensionless parameter ${\Gamma_{\rm g}}/{\Lambda_{\rm g}}$ is given by
\begin{equation}
\frac{\Gamma_{\rm g}}{\Lambda_{\rm g}} = 0.696 {\left( \frac{\rho_{\rm g}}{2 \times 10^{-8}\ {\rm g}\ {\rm cm}^{-3}} \right)} {\left( \frac{T_{\rm g}}{2000\ {\rm K}} \right)}^{- 5/2}.
\end{equation}
Then, we can calculate the equilibrium temperature of chondrules in high-temperature gas as a function of the gas density and the gas temperature, $\rho_{\rm g}$ and $T_{\rm g}$.
Figure \ref{fig5} shows that completely molten droplets turn into supercooled droplets with a temperature of $900\ {\rm K} < T < 1400\ {\rm K}$ when the gas density in the post-shock region is on the order of $\rho_{\rm g} \sim 10^{-8}\ {\rm g}\ {\rm cm}^{-3}$, where $T_{\rm glass} \simeq 900\ {\rm K}$ is the glass transition temperature and $T \lesssim 1400\ {\rm K}$ is the condition for surviving high-speed collisions (Section \ref{sec2.5.2}).
Therefore, the preferred value of the gas density in the pre-shock region, $\rho_{{\rm g}, 0}$, is on the order of $10^{-9}$--$10^{-8}\ {\rm g}\ {\rm cm}^{-3}$ because the gas density increases after the passage of the shock front.
We note, however, that the lower limit of $\rho_{\rm g}$ to maintain the supercooling of chondrule precursors is also dependent on the background temperature (in this study, we simply assume that the background temperature is the same as the pre-shock gas temperature, $T_{0} = 500\ {\rm K}$).
In addition, the effective background temperature may be affected by the optical depth of the chondrule-forming region when the optical depth is close to unity.
We will study the three-dimensional (or axisymmetric two-dimensional) radiative hydrodynamics of planetary bow shocks in the future.

\subsection{Collision frequency}

In Section \ref{sec3.1}, we calculated the velocity evolution of chondrules in the post-shock region.
The velocity depends on the radius of chondrules, and collision of chondrules occurs towing to the difference in the velocity.
Then, we can calculate the collision frequency of chondrules.

\begin{figure}
\centering
\includegraphics[width=\columnwidth]{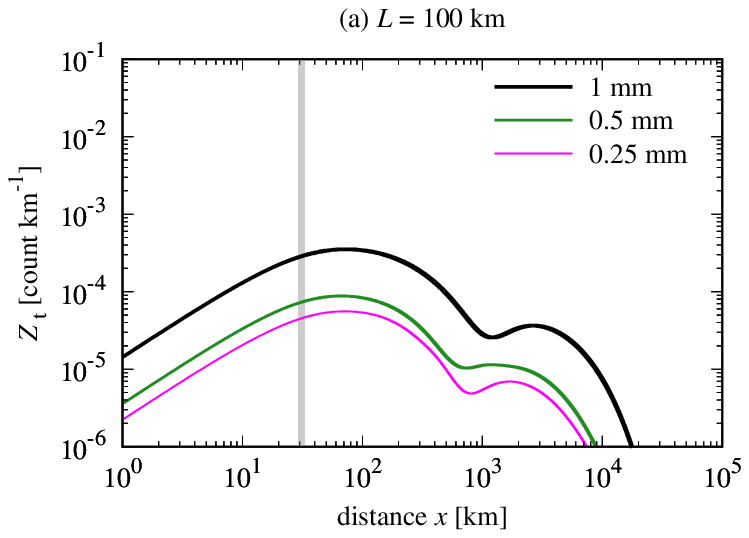}
\includegraphics[width=\columnwidth]{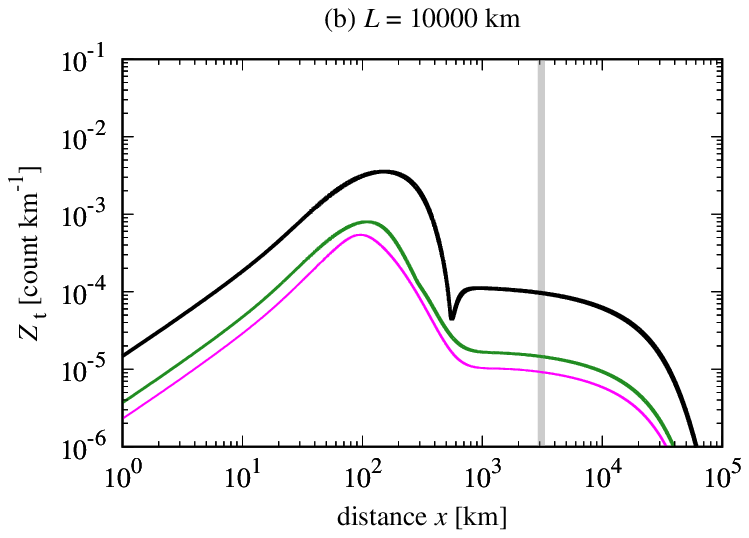}
\caption{
The collision frequency of a target chondrule with any projectile, $Z_{\rm t}$.
(a) For the case of the small-scale shock wave ($L = 100\ {\rm km}$).
(b) For the case of the large-scale shock wave ($L = 10000\ {\rm km}$).
The solid curves represent $Z_{\rm t}$ of chondrules with radii of $r = 1\ {\rm mm}$ (black), $r = 0.5\ {\rm mm}$ (green), and $r = 0.25\ {\rm mm}$ (magenta).
The gray vertical line represents the recondensation line of evaporated fine dust grains.
We assumed that the chondrule mass density in the pre-shock region is $\rho_{{\rm c}, 0} = 6 \times 10^{-12}\ {\rm g}\ {\rm cm}^{-3}$.
}
\label{fig6}
\end{figure}

\begin{figure}
\centering
\includegraphics[width=\columnwidth]{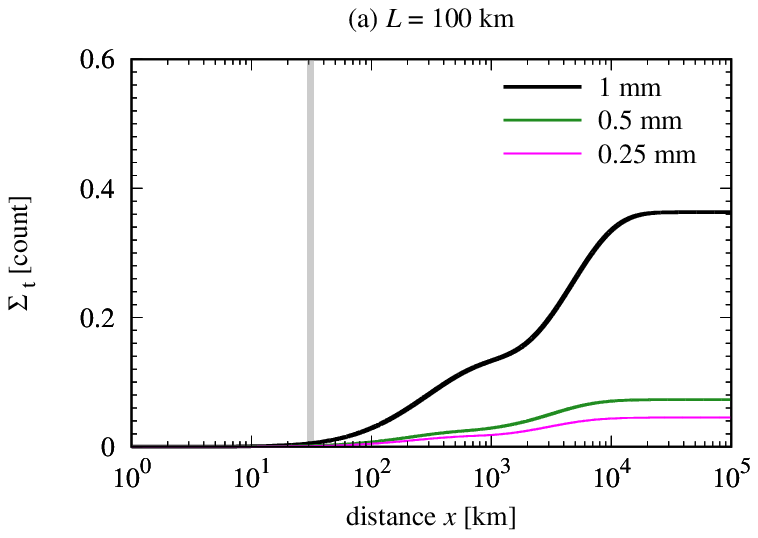}
\includegraphics[width=\columnwidth]{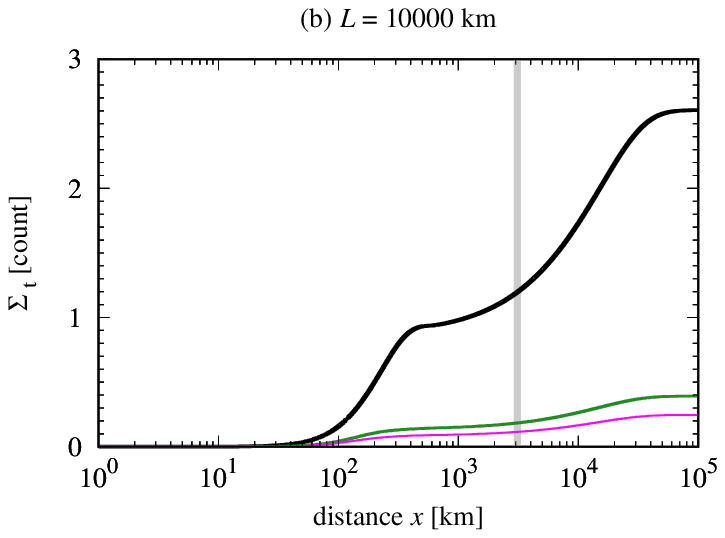}
\caption{
The expected number of collisions for each target chondrule after passing the shock front, $\Sigma_{\rm t}$.
(a) For the case of the small-scale shock wave ($L = 100\ {\rm km}$).
(b) For the case of the large-scale shock wave ($L = 10000\ {\rm km}$).
The solid curves represent $\Sigma_{\rm t}$ of chondrules with radii of $r = 1\ {\rm mm}$ (black), $r = 0.5\ {\rm mm}$ (green), and $r = 0.25\ {\rm mm}$ (magenta).
The gray vertical line represents the recondensation line of evaporated fine dust grains.
}
\label{fig7}
\end{figure}

Figure \ref{fig6} shows the collision frequency of a target chondrule with any projectiles, $Z_{\rm t}$, and Figure \ref{fig7} shows the expected number of collisions for each target chondrule after passing the shock front, $\Sigma_{\rm t}$.
Here, we assumed that the chondrule mass density in the pre-shock region is $\rho_{{\rm c}, 0} = 6 \times 10^{-12}\ {\rm g}\ {\rm cm}^{-3}$.
The chondrule-to-gas mass ratio in the pre-shock region is therefore $\rho_{{\rm c}, 0} / \rho_{{\rm g}, 0} = 2 \times 10^{-3}$, and this value is approximately half of the well-assumed silicate-to-gas mass ratio \citep[$= 4.3 \times 10^{-3}$;][]{Miyake+1993}. 
We can imagine that part of the silicate dust may exist as fine dust grains and others exist as chondrules and/or large dust aggregates.
Therefore, our estimate of $\rho_{{\rm c}, 0} / \rho_{{\rm g}, 0} = 2 \times 10^{-3}$ is reasonable to some extent.

As shown in Figures \ref{fig6} and \ref{fig7}, the collision of chondrules occurs in two stages; the first stage corresponds to where the velocity of chondrules approaches the gas velocity and larger chondrules have larger values of $v$, and the second stage corresponds to where the velocity of chondrules recover to the pre-shock velocity and smaller chondrules have larger values of $v$.
The frequency of collision depends on the spatial scale $L$ if $L$ is comparable to or smaller than the stopping length of chondrules, i.e., $L \lesssim l_{\rm stop}$.
This fact has been previously mentioned by \citet{Jacquet+2014}, and $Z_{\rm t}$ and $\Sigma_{\rm t}$ are small for small-scale shock waves compared with the case of large-scale shock waves.

The expected number of collisions for submillimeter-sized chondrules is lower than unity when we assume $\rho_{{\rm c}, 0} / \rho_{{\rm g}, 0} = 2 \times 10^{-3}$; therefore, most of the chondrule precursors that are heated above their liquidus temperature turn into supercooled droplets and can keep their supercooling state until the recondensation of fine dust grains occurs (see Figure \ref{fig7}).
Conversely, millimeter-sized large chondrules collide frequently, and for the case of large-scale shock waves, most of the millimeter-sized chondrules experience collision when $\rho_{{\rm c}, 0} / \rho_{{\rm g}, 0} \gtrsim 2 \times 10^{-3}$.
After a collision, the supercooled droplet turns into a crystallized chondrule when the collision velocity is below $v_{\rm merge}$, and some of these chondrules have experienced multiple collisions; this is the mechanism of compound chondrule formation \citep[see Figure 2 of][]{Arakawa+2016a}.
We note that the number of collisions $\Sigma_{\rm t}$ is proportional to $\rho_{{\rm c}, 0} / \rho_{{\rm g}, 0}$; then, $\Sigma_{\rm t}$ for submillimeter-sized chondrules could also exceed unity when $\rho_{{\rm c}, 0} / \rho_{{\rm g}, 0} \gtrsim 10^{-2}$.

\begin{figure*}
\centering
\includegraphics[width=0.4\textwidth]{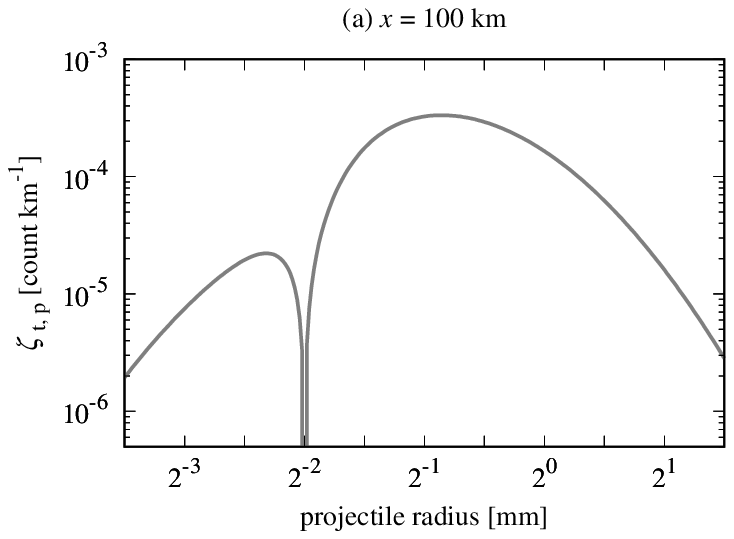}
\includegraphics[width=0.4\textwidth]{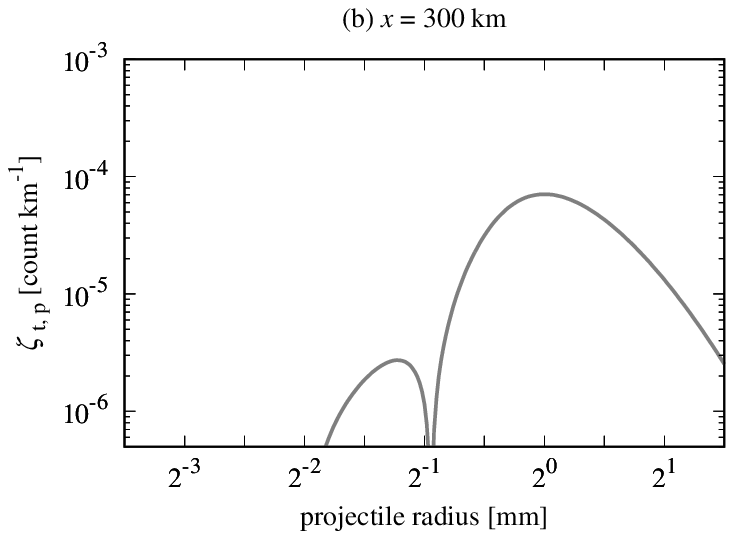}
\includegraphics[width=0.4\textwidth]{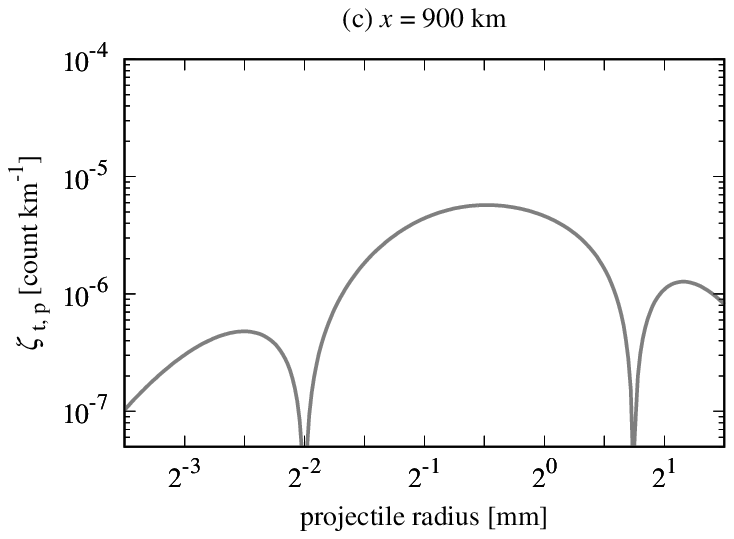}
\includegraphics[width=0.4\textwidth]{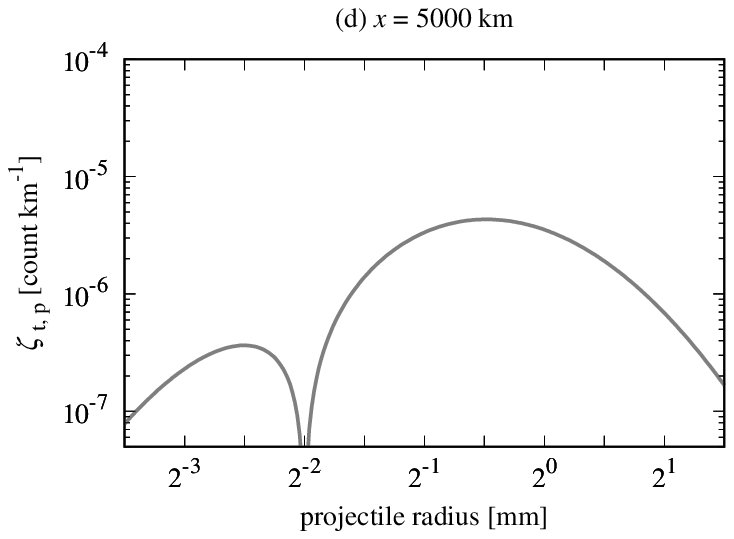}
\includegraphics[width=0.4\textwidth]{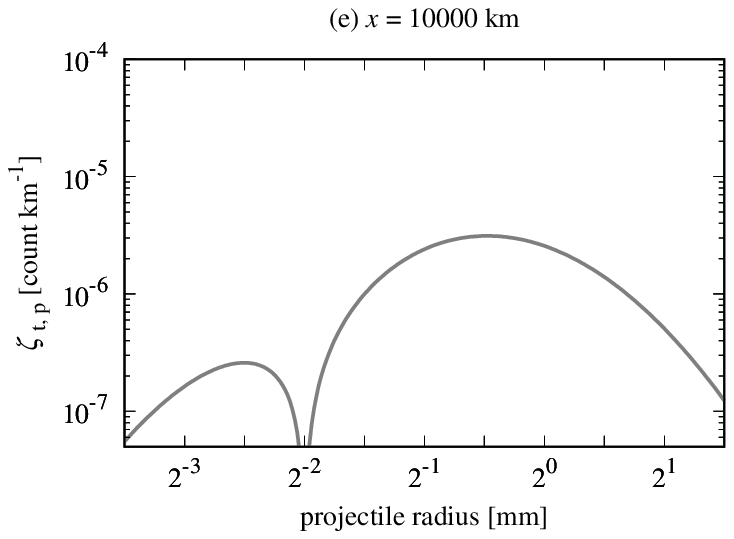}
\includegraphics[width=0.4\textwidth]{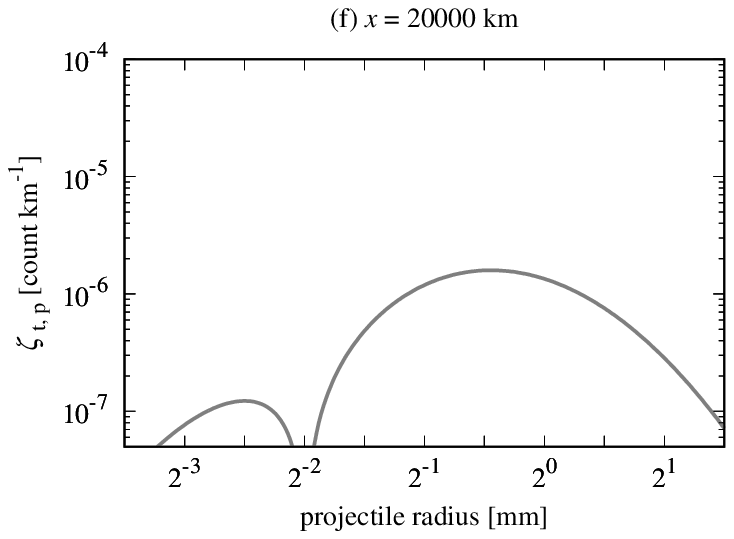}
\caption{
The collision frequency of a target chondrule whose size is $r_{\rm t} = 0.25\ {\rm mm}$ with projectile chondrules whose size is $r_{\rm p}$ at (a) $x = 100\ {\rm km}$, (b) $x = 300\ {\rm km}$, (c) $x = 900\ {\rm km}$, (d) $x = 5000\ {\rm km}$, (e) $x = 10000\ {\rm km}$, and (f) $x = 20000\ {\rm km}$, respectively.
The presented results are for the case of the large-scale shock wave ($L = 10000\ {\rm km}$).
}
\label{fig8}
\end{figure*}

Figure \ref{fig8} shows the collision frequency of a target chondrule whose size is $r_{\rm t} = 0.25\ {\rm mm}$ with projectile chondrules whose size is $r_{\rm p}$.
The peak of the collision frequency distribution is located between $r_{\rm p} = 0.5\ {\rm mm}$ and $1\ {\rm mm}$ for the whole region.
This is due to the balance of the impact velocity, the collisional cross section, and the number density of chondrules; large chondrules have large velocities and large cross sections but small number densities.
As shown in Figure \ref{fig7}, large chondrules tend to crystallize earlier, and small ones tend to be secondaries.
In addition, the peak of the size-frequency distribution is located around $r \sim 0.25\ {\rm mm}$.
Therefore, the radius of secondaries $r_{\rm sec}$ may be distributed around $r_{\rm sec} \sim 0.25\ {\rm mm}$, and the typical radius of primaries $r_{\rm pri}$ may be $r_{\rm pri} \sim 0.5$--$1\ {\rm mm}$.

The secondary-to-primary size ratio, $\Delta_{\rm sp} \equiv r_{\rm sec} / r_{\rm pri}$, has been measured in thin sections by a few studies \citep[e.g.,][]{Wasson+1995}, and the mean value of $\Delta_{\rm sp}$ for compound chondrules in ordinary chondrites is $\sim 0.3$.
This value seems to be consistent with the calculated collision frequency distribution shown in Figure \ref{fig8} (see also Figure \ref{fig9}); although the observation in the thin section is somewhat biased and the real value of $\Delta_{\rm sp}$ may be somewhat larger than $0.3$ \citep[see][]{Ciesla+2004b}.

\subsection{Collisions of supercooled droplets}

In our calculation, we obtain the temperature and the velocity of chondrules simultaneously.
Therefore, we can compare the impact velocity of chondrules with different radius $v_{\rm imp}$ and the critical velocity for collisional merging $v_{\rm merge}$ and sticking $v_{\rm stick}$, which are dependent on the temperature of chondrules.
Hereafter, we focus on the case of the large-scale shock wave ($L = 10000\ {\rm km}$).

\begin{figure*}
\centering
\includegraphics[width=0.32\textwidth]{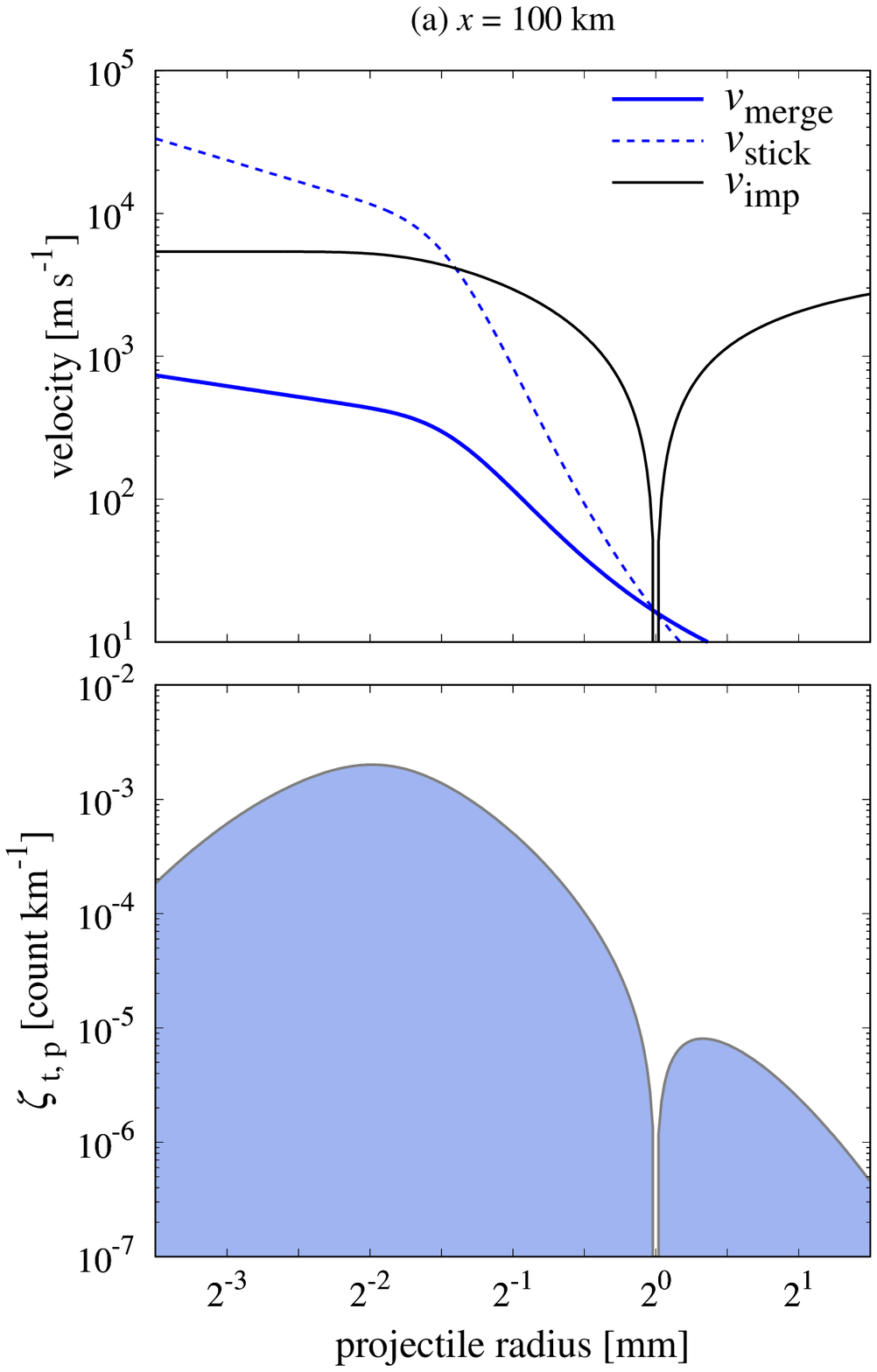}
\includegraphics[width=0.32\textwidth]{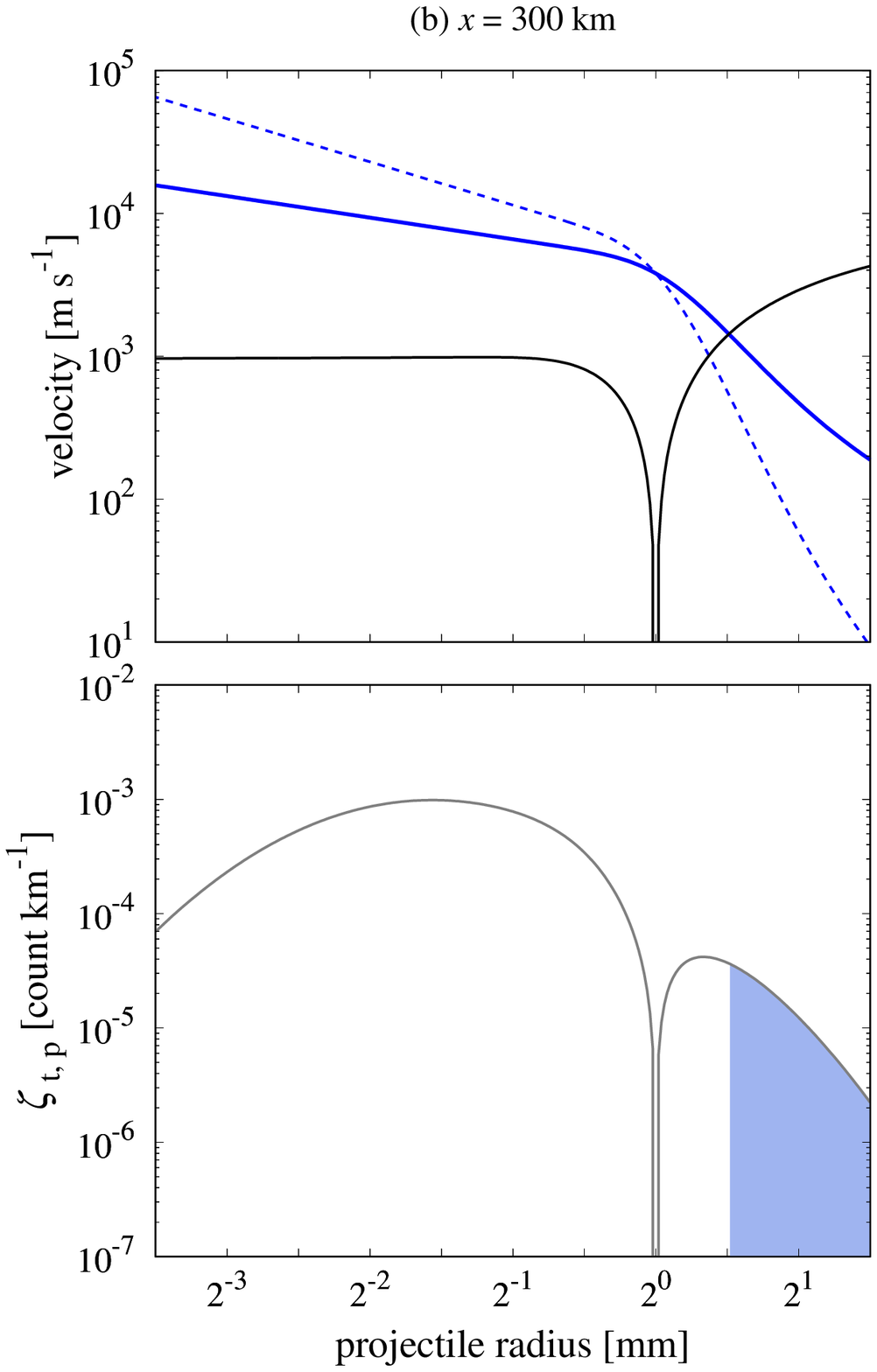}
\includegraphics[width=0.32\textwidth]{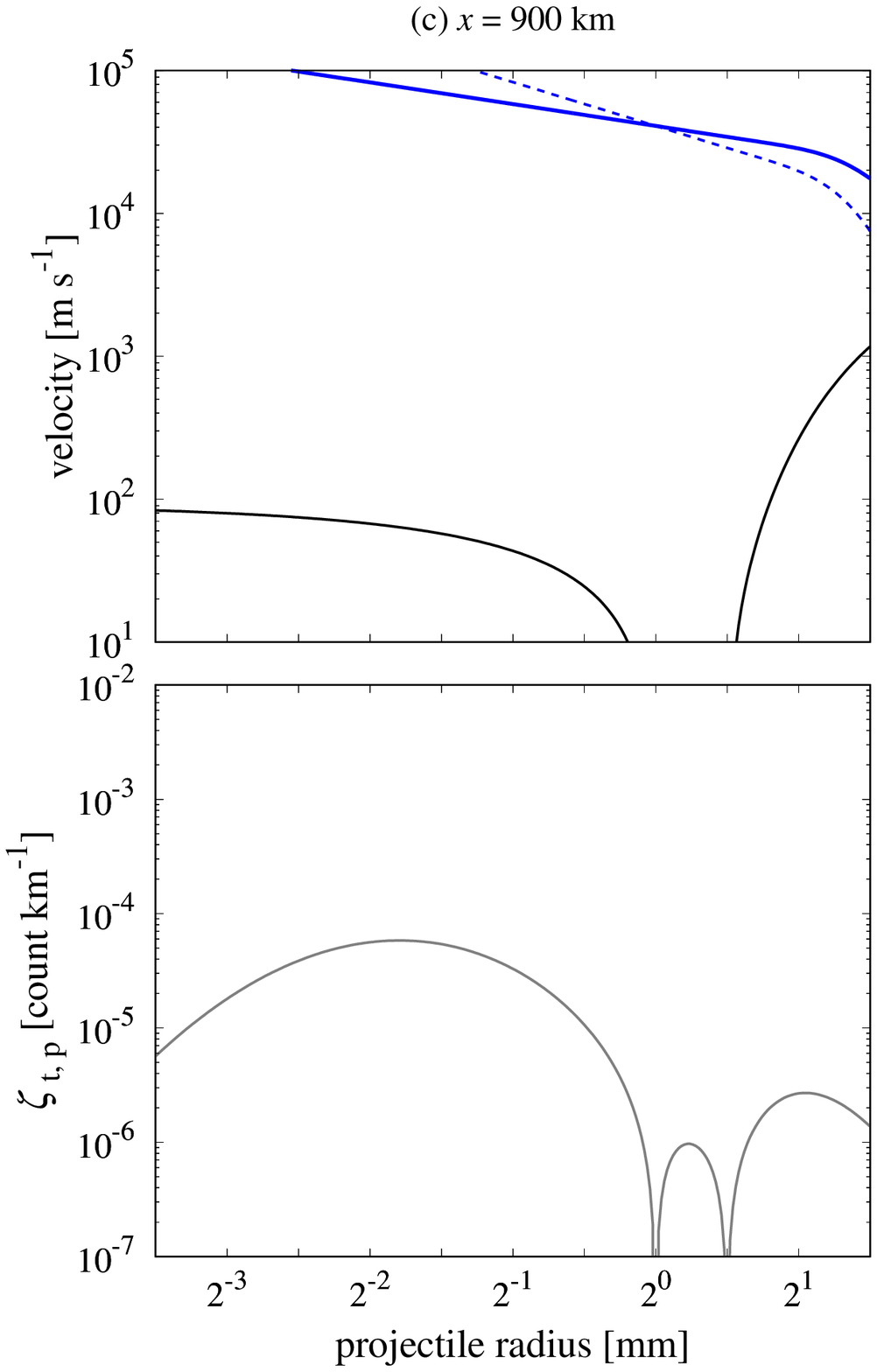}
\caption{
{\it Upper panels}: the critical velocity for collisional merging $v_{\rm merge}$, sticking $v_{\rm stick}$, and the impact velocity $v_{\rm imp}$.
{\it Lower panels}: the collision frequency of a target chondrule whose size is $r_{\rm t} = 1\ {\rm mm}$ with projectile chondrules ${\zeta_{\rm t,p} (\varnothing_{\rm t} = - 1, \varnothing_{\rm p}, x)}$.
The shaded regions show where non-crystallized targets would disrupt when they collide with projectiles, i.e., $v_{\rm imp} > v_{\rm merge}$.
The presented results are for the case of the large-scale shock wave ($L = 10000\ {\rm km}$).
(a) The snapshot at $x = 100\ {\rm km}$.
(b) The snapshot at $x = 300\ {\rm km}$.
(c) The snapshot at $x = 900\ {\rm km}$.
}
\label{fig9}
\end{figure*}

Figure \ref{fig9} shows the critical velocity for collisional merging $v_{\rm merge}$, sticking $v_{\rm stick}$, and the impact velocity $v_{\rm imp}$ for target chondrules with $r_{\rm t} = 1\ {\rm mm}$.
From Figure \ref{fig4}a, the temperature of chondrules with $r_{\rm t} = 1\ {\rm mm}$ rapidly decreases before the distance from the shock front reaches $x \simeq 300\ {\rm km}$.
The critical velocities $v_{\rm merge}$ and $v_{\rm stick}$ are strongly dependent on the temperature of the target and projectile chondrules (see Figure \ref{fig2}).
Therefore, both $v_{\rm merge}$ and $v_{\rm stick}$ significantly increase before the distance from the shock front reaches $x \simeq 300\ {\rm km}$.
In addition, the impact velocity $v_{\rm imp}$ falls below $1\ {\rm km}\ {\rm s}^{-1}$ for $x \gtrsim 200$--$300\ {\rm km}$; then, $v_{\rm merge}$ and $v_{\rm stick}$ overcome $v_{\rm imp}$.

As a conclusion, compound chondrules with a primary radius of $\sim 1\ {\rm mm}$ would be formed via collisions of supercooled droplets in the post-shock region where the distance from the shock front exceeds $x \gtrsim 300\ {\rm km}$, although the suitable location for compound chondrule formation must depend on the detailed characteristics of the specific chondrule-forming shock waves.

The lower panels of Figure \ref{fig9} show the ${\zeta_{\rm t,p} (\varnothing_{\rm t}, \varnothing_{\rm p}, x)}$ of supercooled chondrules with a target radius of $r_{\rm t} = 1\ {\rm mm}$, i.e., $\varnothing_{\rm t} = - 1$.
The size-frequency distribution of projectiles is a maximum at $r_{\rm p} \sim 0.25$--$0.5\ {\rm mm}$.
As shown in Figure \ref{fig7}, the expected number of collisions $\Sigma_{\rm t}$ is lower for smaller chondrules.
Then, the probability that the small projectile chondrule is supercooled while the large target chondrule is already crystallized is higher than the probability that the small projectile chondrule is crystallized while the large target chondrule is still in the supercooled state.
Therefore, compound chondrules whose secondary-to-primary size ratio is $\Delta_{\rm sp} \sim 0.3$ may be formed via a collision between crystallized and supercooled chondrules in the post-shock region, as already mentioned (see Figure \ref{fig8}).

Here, we note that some of the collisions must cause the splashing of supercooled droplets when they collide with high speed and/or high temperature, although the fraction of disruption is lower than unity when we assume $\rho_{{\rm c}, 0} / \rho_{{\rm g}, 0} \sim 2 \times 10^{-3}$.
\citet{Jacquet+2014} noted that chondrules can also be destroyed by continuous erosion through the collisions of fragments produced by other catastrophic collision events.
In this study, we do not take into consideration this ``sandblasting'' effect, however.
Whether the collisions of fragments would be critical or not is dependent on the size distribution of fragments, and future studies on this point are needed.

\subsection{Survivability of crystallized chondrules}
\label{sec3.5}

The evaporated fine dust grains would recondense when the gas temperature decreases below the dust condensation temperature $T_{\rm c}$ (we assumed $T_{\rm c} = 1600\ {\rm K}$ in this study).
The location of the dust condensation line $x_{\rm c}$ is therefore $x_{\rm c} \sim 0.3 L$ when we assume the gas temperature is determined by Equation (\ref{eqTg}).
After the recondensation of fine dust grains, supercooled droplets are crystallized by the accretion of condensates onto chondrule precursors \citep[e.g.,][]{Nagashima+2006,Nagashima+2008}.

\begin{figure*}
\centering
\includegraphics[width=0.32\textwidth]{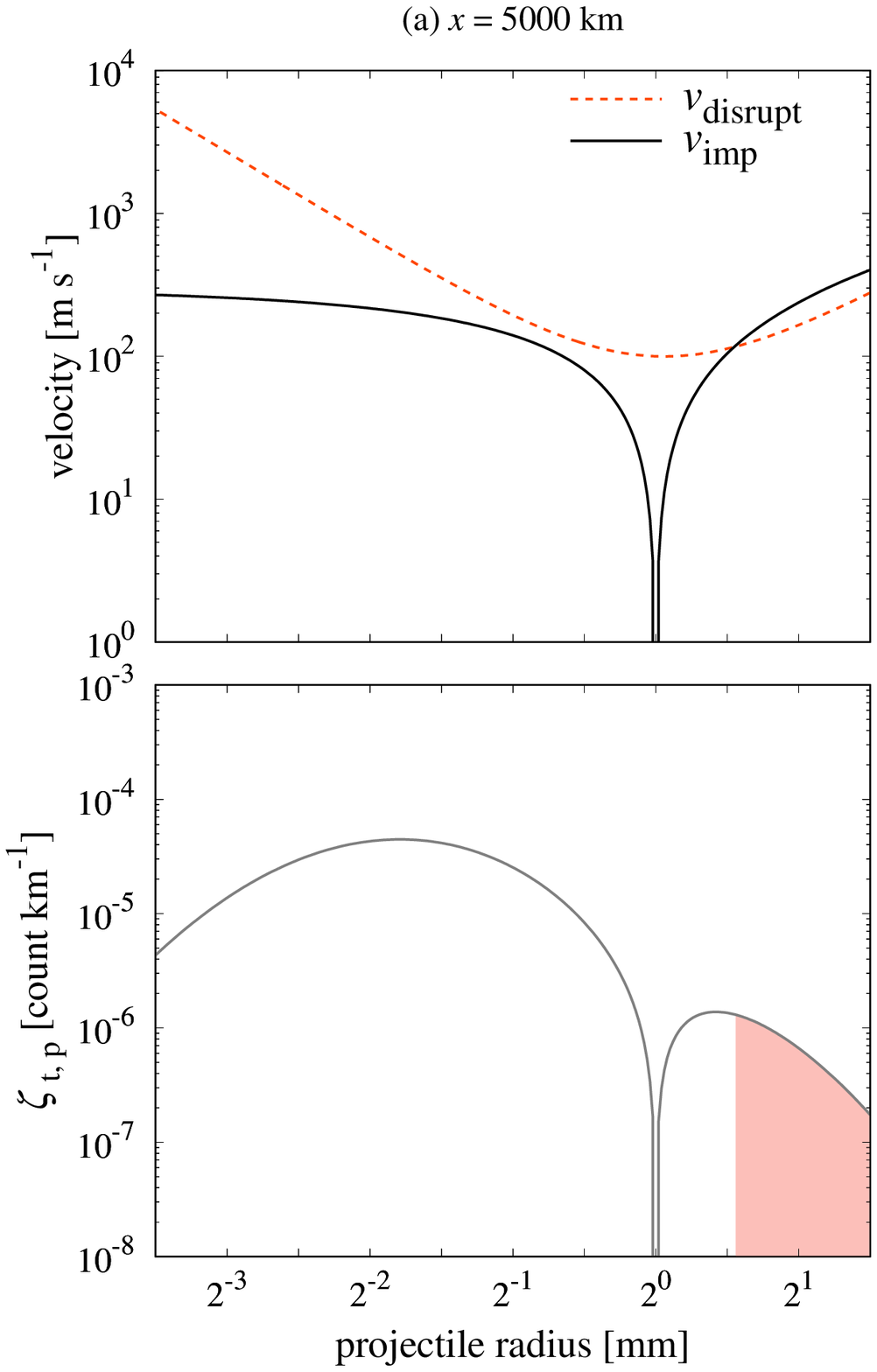}
\includegraphics[width=0.32\textwidth]{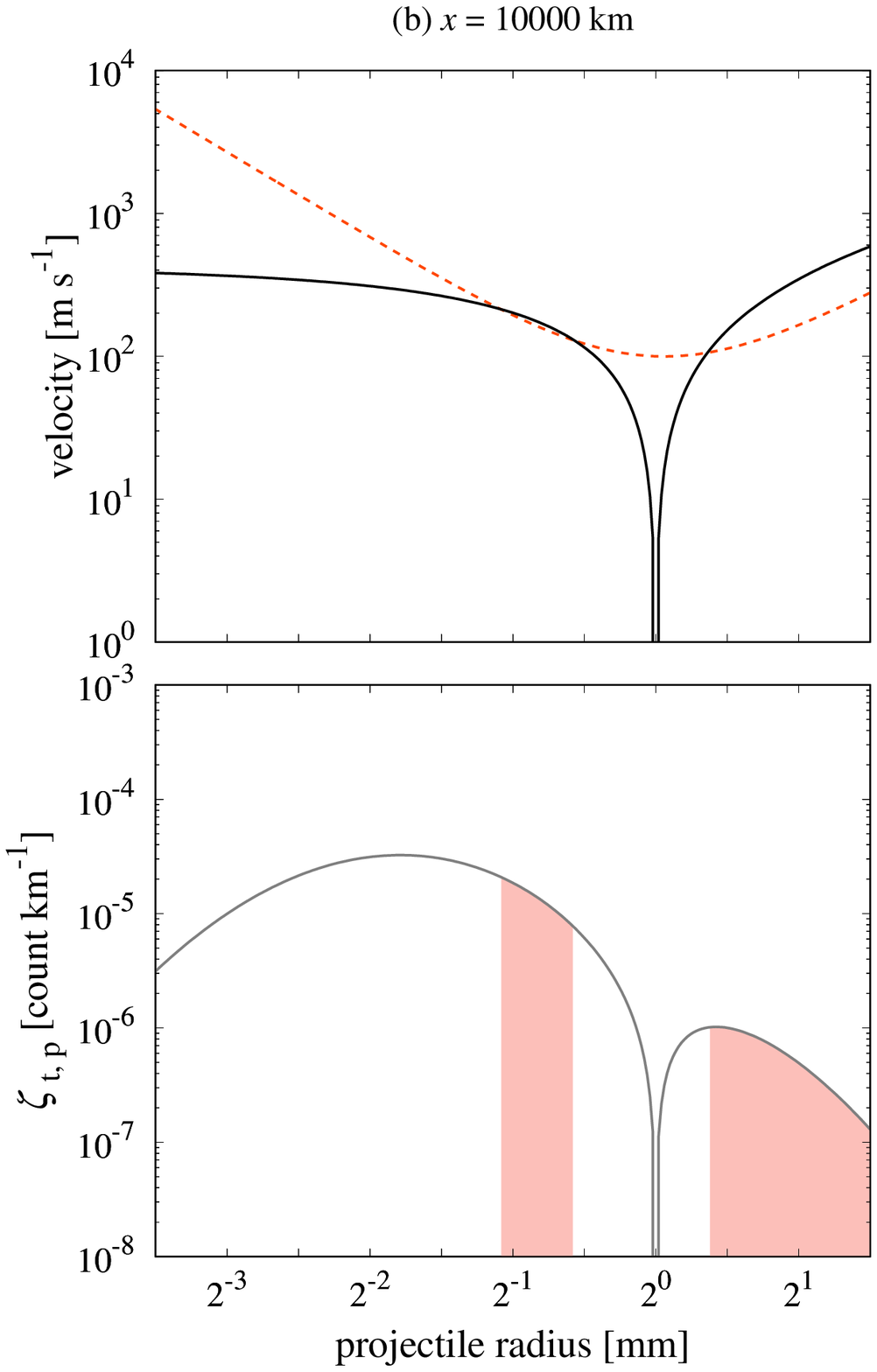}
\includegraphics[width=0.32\textwidth]{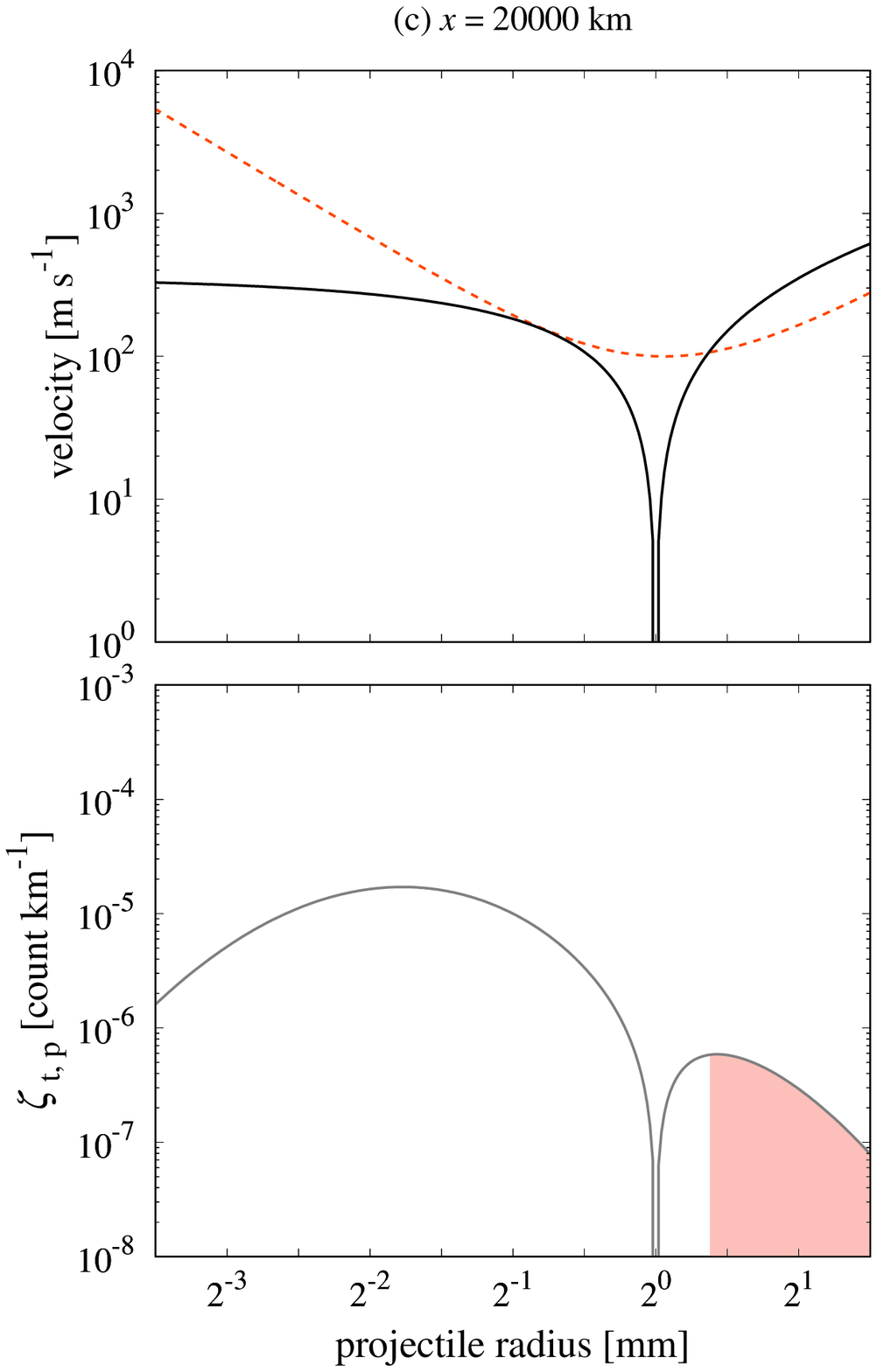}
\caption{
{\it Upper panels}: the critical velocity for catastrophic disruption $v_{\rm disrupt}$ and the impact velocity $v_{\rm imp}$.
{\it Lower panels}: the collision frequency of a target chondrule whose size is $r_{\rm t} = 1\ {\rm mm}$ with projectile chondrules ${\zeta_{\rm t,p} (\varnothing_{\rm t} = - 1, \varnothing_{\rm p}, x)}$.
The shaded regions show where crystallized targets would disrupt when they collide with projectiles, i.e., $v_{\rm imp} > v_{\rm disrupt}$.
The presented results are for the case of the large-scale shock wave ($L = 10000\ {\rm km}$).
(a) The snapshot at $x = 5000\ {\rm km}$.
(b) The snapshot at $x = 10000\ {\rm km}$.
(c) The snapshot at $x = 20000\ {\rm km}$.
}
\label{fig10}
\end{figure*}

Here, we investigate whether crystallized chondrules can avoid catastrophic disruption after their crystallization.
We compare $v_{\rm imp}$ and $v_{\rm disrupt}$; then, the survivability of crystallized chondrules is evaluated.
The upper panels of Figure \ref{fig10} show the critical velocity for catastrophic disruption $v_{\rm disrupt}$ and the impact velocity $v_{\rm imp}$ for target chondrules with $r_{\rm t} = 1\ {\rm mm}$, and the lower panels show the ${\zeta_{\rm t,p} (\varnothing_{\rm t}, \varnothing_{\rm p}, x)}$ of supercooled chondrules with a target radius of $r_{\rm t} = 1\ {\rm mm}$.

Without performing numerical simulations, we can roughly evaluate the impact velocity of chondrules by simple analytical calculations.
The impact velocity of large and small chondrules is approximately given by the relative velocity of the large chondrule from the gas.
The velocity of chondrules with respect to the shock front is given by the following time differential equation:
\begin{equation}
\frac{{\rm d}v}{{\rm d}t} = - \frac{3}{4} \frac{C_{\rm D}}{2} \frac{\rho_{\rm g}}{\rho} \frac{{\left| v - v_{\rm g} \right|} {\left( v - v_{\rm g} \right)}}{r} \simeq - 3 c_{\rm s} \frac{\rho_{\rm g}}{\rho} \frac{{\left( v - v_{\rm g} \right)}}{r},
\label{eqdvdx1}
\end{equation}
where $t$ is the time and we assume $C_{\rm D} \sim 8 / s$.
For the case of large-scale shock waves, the relative velocity of the chondrule from the gas is significantly smaller than the gas velocity, i.e., ${\left| v - v_{\rm g} \right|} \ll v_{\rm g}$.
Then, the differential of ${\left| v - v_{\rm g} \right|}$ is also negligible, i.e., $\left| {{\rm d}{\left( v - v_{\rm g} \right)}} / {{\rm d}t} \right| \ll {\left| {{\rm d}v_{\rm g}} / {{\rm d}t} \right|}$.
This means that the differential of the velocity of chondrules is approximately given by the gas velocity and the spatial scale of the shock wave as follows:
\begin{equation}
{\left| \frac{{\rm d}v}{{\rm d}t} \right|} \simeq {\left| \frac{{\rm d}v_{\rm g}}{{\rm d}t} \right|} \sim \frac{{\left( v_{0} - v_{\rm post} \right)}^{2}}{L}.
\label{eqdvdx2}
\end{equation}
Therefore, from the combination of Equations (\ref{eqdvdx1}) and (\ref{eqdvdx2}), the relative velocity of the chondrule from the gas can be evaluated as follows:
\begin{eqnarray}
{\left| v - v_{\rm g} \right|} &\simeq& \frac{1}{3} \frac{\rho}{\rho_{\rm g}} \frac{r}{L} \frac{{\left( v_{0} - v_{\rm post} \right)}^{2}}{c_{\rm s}} \nonumber \\
                               &\sim& 3 \times 10^{2}\ {\left( \frac{\rho_{\rm g}}{10^{-8}\ {\rm g}\ {\rm cm}^{-3}} \right)}^{-1} {\left( \frac{L}{10^{4}\ {\rm km}} \right)}^{-1} \nonumber \\
                               && \cdot {\left( \frac{r}{1\ {\rm mm}} \right)}\ {\rm m}\ {\rm s}^{-1}.
\end{eqnarray}
Our numerical simulations also confirmed that the typical impact velocity of chondrules of $1\ {\rm mm}$ in radius is approximately $300\ {\rm m}\ {\rm s}^{-1}$, and disruptive collisions are minor among all collisions (see Figure \ref{fig10}) when the spatial scale is $L \gtrsim 10000\ {\rm km}$.
The expected fraction for catastrophic disruption is, therefore, lower than unity for chondrules whose radius is less than $1\ {\rm mm}$ in this case.
We note that, when the spatial scale is smaller than $100\ {\rm km}$ (i.e., $L \ll l_{\rm stop}$), the expected number of collisions itself is far lower than unity, and neither do we need to consider the catastrophic disruption of chondrules, although it depends on the chondrule-to-gas mass ratio.

The impact velocity is inversely proportional to the spatial scale of the shock $L$, and the necessary condition for chondrule survival may be $L \gtrsim 10000\ {\rm km}$.
When the shock waves are caused by eccentric planetary bodies, the spatial scale of the shock wave $L$ is approximately a few to ten times larger than the planetary radius ${\cal R}_{\rm p}$ \citep[e.g.,][]{Morris+2012,Boley+2013}, although $L / {\cal R}_{\rm p}$ depends on the opacity, the shock velocity, and so on.
Therefore, planetary bow shocks caused by $1000\ {\rm km}$-sized protoplanets may be potent candidates for the chondrule formation mechanism from the point of view of chondrule survivability.

\section{Discussion}

\subsection{Chondrule-to-gas mass ratio}

It is usually assumed that the silicate-to-gas mass ratio is approximately $4.3 \times 10^{-3}$ \citep{Miyake+1993}, and part of the silicate dust can exist as fine dust grains, while others formed chondrules and/or much larger dust aggregates.
Therefore, we assume that the chondrule-to-gas mass ratio in the pre-shock region is $2 \times 10^{-3}$.
However, when chondrules sediment at the midplane of the solar nebula, the chondrule-to-gas mass ratio at the midplane $\rho_{\rm c} / \rho_{\rm g}$ becomes significantly higher than the chondrule-to-gas surface density ratio $\chi$.
Here, we evaluate whether the sedimentation of chondrules would occur.

When the radius of a chondrule is smaller than the mean-free path of gas molecules, i.e., the gas drag force on the chondrule is determined by Epstein's law, the dimensionless stopping time called the Stokes number ${\rm St}$ is given by
\begin{eqnarray}
{\rm St} &=& \sqrt{\frac{\pi}{8}} \frac{\rho r \Omega_{\rm K}}{\rho_{\rm g} c_{\rm s}} \nonumber \\
         &\sim& 7 \times 10^{-5}\ {\left( \frac{r}{1\ {\rm mm}} \right)}\ {\left( \frac{\rho_{\rm g}}{3 \times 10^{-9}\ {\rm g}\ {\rm cm}^{-3}} \right)}^{-1}\ \nonumber \\
         && \cdot {\left( \frac{T_{0}}{500\ {\rm K}} \right)}^{-1/2}\ {\left( \frac{R}{1\ {\rm au}} \right)}^{-3/2},
\end{eqnarray}
where $\Omega_{\rm K}$ is the Kepler frequency and $R$ is the distance from the sun \citep{Weidenschilling1977}.
The gas scale height $h_{\rm g}$ is given by $h_{\rm g} = c_{\rm s} / \Omega_{\rm K}$, and the chondrule scale height $h_{\rm c}$ is given by \citep{Youdin+2007}:
\begin{equation}
\frac{h_{\rm c}}{h_{\rm g}} = {\left( 1 + \frac{{\rm St}}{\alpha_{\rm t}} \frac{1 + 2 {\rm St}}{1 + {\rm St}} \right)}^{-1/2},
\end{equation}
where $\alpha_{\rm t}$ is a dimensionless turbulent parameter.
Then, the chondrule-to-gas mass ratio at the midplane is given by $\rho_{\rm c} / \rho_{\rm g} = {\left( {h_{\rm c}} / {h_{\rm g}} \right)}^{-1} \chi$.

The value of the dimensionless parameter $\alpha_{\rm t}$ for our solar nebula is unclear; however, some protoplanetary disks (e.g., the disk around HL Tau) have a turbulent viscosity that is equivalent to $\alpha_{\rm t}$ in the range of $10^{-4}$ to $10^{-3}$ \citep[e.g.,][]{Pinte+2016,Okuzumi+2016} in the outer regions.
Conversely, the dimensionless parameter $\alpha_{\rm t}$ for the inner region of the disk has not yet been revealed by astronomical observations.
Theoretical studies suggest that $\alpha_{\rm t}$ is up to $10^{-3}$ or higher when the magneto-rotational instability is active, while $\alpha_{\rm t}$ may be on the order of $10^{-4}$ if the magneto-rotational instability is inactive \citep[e.g.,][]{Balbus+1991}.
Therefore, the scale heights of gas and chondrules, $h_{\rm g}$ and $h_{\rm c}$, should be almost the same when the gas density at the midplane is $\rho_{\rm g} \sim 3 \times 10^{-9}\ {\rm g}\ {\rm cm}^{-3}$.
In this case, $\rho_{\rm c} / \rho_{\rm g}$ is approximately given by $\rho_{\rm c} / \rho_{\rm g} \simeq \chi$ and we do not need to consider the enrichment of chondrules at the disk midplane.

\subsection{Location of the chondrule-forming region}

We give a constraint on the location of the chondrule formation from the point of view of the gravitational stability of the solar nebula.
The stability of the disk is measured by Toomre's $Q$ value, defined by \citep{Toomre1964},
\begin{eqnarray}
Q &=& \frac{c_{\rm s} \Omega_{\rm K}}{\pi G \cdot {\left( \sqrt{2 \pi} h_{\rm g} \rho_{\rm g} \right)}} \nonumber \\
  &\simeq& 25\ {\left( \frac{R}{1\ {\rm au}} \right)}^{-3}\ {\left( \frac{\rho_{\rm g}}{3 \times 10^{-9}\ {\rm g}\ {\rm cm}^{-3}} \right)},
\end{eqnarray}
and the gas disk becomes unstable when $Q \lesssim 2$ and the above equation gives the upper limit of the gas density.

As shown in Figure \ref{fig5}, the favored gas density to keep molten chondrules in the supercooled state is $\rho_{\rm g} \sim 10^{-8}\ {\rm g}\ {\rm cm}^{-3}$ in the post-shock region.
This value corresponds to $\rho_{{\rm g}, 0} \sim 10^{-9}$--$10^{-8}\ {\rm g}\ {\rm cm}^{-3}$ in the pre-shock region.
Then, the location of the chondrule-forming region may be within a few astronomical units from the sun if chondrules are formed by optically thin shock waves.
This region overlaps with the location of the inner part of the asteroid belt, which is mostly dominated by S-type asteroids \citep[e.g.,][]{DeMeo+2014}, and this coincidence may indicate that chondrules in ordinary chondrites are formed via shock-wave heating in the inner solar nebula because S-type asteroids are the parent bodies of ordinary chondrites \citep{Nakamura+2011}, while chondrules in carbonaceous chondrites may be linked to different events and locations.

\subsection{Volatile retention}

Chondrules contain volatile elements such as sodium, potassium, and sulfur in their interiors.
This implies that chondrules are formed by flash-heating/rapid-cooling events \citep[e.g.,][]{Tachibana+2005,Rubin2010,Wasson2012} or the ambient environments where chondrules melted under a high partial pressure of lithophile elements \citep[e.g.,][]{Alexander+2008,Fedkin+2013}.
The latter hypothesis, called ``dust enrichment'', originates from the assumption that porphyritic chondrules, which are the main type among all chondrules, may be formed with a low cooling rate \citep[$\sim 10^{-3}$--$1\ {\rm K}\ {\rm s}^{-1}$;][and references therein]{Desch+2012}.
This assumption originates from the results of classical furnace-based crystallization experiments \citep[e.g.,][]{Radomsky+1990}; however, several estimations based on some chondrule features, such as overgrowth thicknesses on relict grains \citep[e.g.,][]{Wasson+2003} and rim formation for barred olivine chondrules \citep{Miura+2010b}, give much higher cooling rates \citep[$\sim 200$--$2000\ {\rm K}\ {\rm s}^{-1}$;][]{Miura+2014}.
Moreover, porphyritic textures may be reproduced by multiple melting processes \citep[e.g.,][]{Rubin2010} and they can also be formed via supercooled precursors \citep[e.g.,][]{Srivastava+2010,Seto+2017}.
Therefore, dust enrichment is not necessarily needed for volatile retainment when the heating/cooling rates around their liquidus temperature are high enough.

In addition, chondrules in different chondrite groups have different average sizes \citep[e.g.,][]{Scott2007}, and chondrite groups with large average chondrule sizes (e.g., CV chondrites) tend to have less bulk sodium than groups with small average chondrule sizes \citep{Wasson+1988} and low proportions of nonporphyritic chondrules \citep{Rubin2010}.
These features can be interpreted as a result of multiple flash-melting events \citep[e.g.,][]{Rubin2010}, and the constraint on the cooling rate can be mitigated.

\subsection{Metal grains}

Recently, \citet{Libourel+2018} found a notable absence of metal grains in barred olivine chondrules.
The absence of metal grains in completely molten chondrule precursors was theoretically predicted by \citet{Uesugi+2005,Uesugi+2008}.
In addition, the unique occurrence of metal grains in the core region of magnesium-rich olivine crystals of porphyritic chondrules suggests that the metal grains act as seeding agents during the crystal growth of the olivine crystals in porphyritic chondrules \citep{Libourel+2018}, and the difference in the textures of porphyritic or barred olivine chondrules is linked to the presence/absence of iron-nickel metal grains.

After the ejection of metal grains from molten chondrule precursors, metal grains may collide and merge with other metal grains.
\citet{Okabayashi+2019} measured the abundances of highly siderophile elements on metal grains from type 3 ordinary chondrites and found that larger metal grains have relatively homogeneous abundances of highly siderophile elements that are close to the bulk metal composition.
This observed trend is consistent with the idea that some of the metal grains collided and merged with other metal grains \citep{Okabayashi+2019}.
For iron and nickel, \citet{Leliwa-Kopystynski+1984} performed collision experiments by using $8\ {\rm mm}$-sized projectiles.
The critical velocity for collisional sticking is $v_{\rm stick} \sim 500\ {\rm m}\ {\rm s}^{-1}$ when the temperature is $290\ {\rm K}$, and the estimated $v_{\rm stick}$ at $1800\ {\rm K}$ is approximately $300\ {\rm m}\ {\rm s}^{-1}$.
Therefore, collisional sticking of ejected metal grains could occur in the post-shock region.

\subsection{Early formation of Jupiter}

If chondrules are formed by bow shocks caused by eccentric planetary bodies, the existence of both Jupiter and the nebular gas in the chondrule-forming era is a necessary condition.
Although the onset of chondrule formation is still debated \citep[e.g.,][]{Kita+2012,Bollard+2017,Pape+2019}, both lead-lead ages and aluminum-magnesium ages show that the onset of chondrule formation is approximately 2 million years after the formation of calcium-aluminum-rich inclusions, or much earlier.
Therefore, Jupiter must be formed within 2 million years in the solar nebula.
We note that the early formation of proto-Jupiter is also favored in the context of the chemical dichotomy between carbonaceous and non-carbonaceous meteorite groups \citep[e.g.,][]{Kruijer+2017} and the preservation of calcium-aluminum-rich inclusions in the carbonaceous chondrite formation region \citep{Desch+2018}.

\subsection{Accretion of chondrules}

There are many studies of the accretion process of chondrules, and some of these studies focus on the effect of fine dust grains accreted onto chondrules.
It is known that some of the chondrules in ordinary and carbonaceous chondrites are rimmed by fine dust grains \citep[$\sim 15\%$ for chondrules in Allende CV3 chondrite,][]{Simon+2018}.
Theoretical studies have also revealed that free-floating chondrules in a protoplanetary disk can obtain porous dust layers \citep[e.g.,][]{Xiang+2019}, which help dust-rimmed chondrules stick together when they collide \citep{Beitz+2012,Gunkelmann+2017}.

Evaporation and recondensation by shock-wave heating events change the size-frequency distribution of fine dust grains \citep[e.g.,][]{Miura+2010a}.
When the cooling rate of evaporated dust is large, the condensates could be nanograins, which would be beneficial for the direct aggregation of silicate dust aggregates \citep{Arakawa+2016b}.
However, when fluffy aggregates constituted by chondrules and fine dust grains collide at large velocities, the chondrules in fluffy matrices may be ejected to the solar nebula again \citep{Arakawa2017}.
Then, the growth of dust-rimmed chondrules may be impeded when they reach a few centimeters in radius.

Meanwhile, these centimeter-sized aggregates have the potential to turn into planetesimals via the streaming instability driven by differences in the motions of the gas and dust particles in the disk \citep{Carrera+2015,Yang+2017}.
In addition, the typical radius of planetesimals formed via the streaming instability is $\sim 10^{2}\ {\rm km}$ \citep{Simon+2016}, which is roughly consistent with the estimated radius of the ordinary chondrite parent bodies \citep[e.g.,][]{Henke+2012a,Henke+2012b}.

The other idea is that chondrules accrete onto planetesimals that already exist in the gaseous solar nebula \citep[e.g.,][]{Hasegawa+2016b,Matsumoto+2017}.
\citet{Matsumoto+2017} calculated the chondrule accretion onto a protoplanet and planetesimals in the oligarchic growth stage \citep[e.g.,][]{Kokubo+1998} and found that approximately half of the chondrules accrete onto the protoplanet, while the other half accrete onto planetesimals with an accretion timescale of $\sim 10^{6}$ years.
In this case, some of the chondrules should have stayed in the solar nebula for a few million years; this timescale is consistent with the fact that some of the chondrules have experienced multiple melting events with the time interval of $\sim 10^{6}$ years \citep[e.g.,][]{Akaki+2007}.

Planetary bodies with moderate eccentricities ($e_{\rm p} \sim 10^{-2}$--$10^{-1}$) accrete chondrule-sized particles more efficiently than planetary bodies in circular orbits; however, the accretion efficiency drops drastically when the eccentricity becomes far larger than $10^{-1}$ \citep{Liu+2018}.
Therefore, it may be difficult to grow eccentric planetesimals/protoplanets into terrestrial planets when they have a large eccentricity.
The excitation of eccentricity increases the gas drag; then, the eccentricity and semimajor axis are quickly damped around $1\ {\rm au}$ \citep{Nagasawa+2014,Nagasawa+2019}, although the location is dependent on the physical properties of the disk.
The migration of planetesimals may cause the concentration of circular planetesimals around $R \sim 1\ {\rm au}$.
This concentration of planetesimals could have the potential to explain why two large terrestrial planets, Venus and Earth, formed at approximately $1\ {\rm au}$ \citep[e.g.,][]{Hansen2009,Walsh+2016}.

\section{Conclusion}

We explored the possibility that compound chondrules are formed via the collisions of supercooled precursors in shock waves.
The shock-wave heating model is one of the prime candidates to explain the origin of chondrules. However, there is one challenge to this model:
chondrule precursors of different sizes must have different velocities in the post-shock region and they should collide with high speed (approximately a few ${\rm km}\ {\rm s}^{-1}$), which may lead to their destruction upon collision rather than compound chondrule formation if they were completely molten.

As it is, \citet{Arakawa+2016a} revealed that compound chondrules may be formed via collisions of supercooled precursors.
Supercooling is the state where liquids do not solidify even below their solidus temperature.
Supercooled chondrule precursors have large viscosity, and their critical velocity for collisional sticking is higher than that of completely molten precursors.
Therefore, the destruction of chondrules could be avoided when we consider the supercooling of chondrule precursors.

We calculated the velocity and the temperature of chondrule precursors in optically thin shock waves.
We found that, in optically thin shock waves, chondrule precursors can maintain their supercooling until the fine dust grains condense and supercooled precursors crystallize via accretion of fine dust grains.
As a first step toward more comprehensive modeling, we considered one-dimensional normal shocks and we assumed a simple gas structure; subsequently, the dynamics of chondrules was simulated in the given gas flow.

Our key findings are summarized as follows.
\begin{enumerate}
  \item Because supercooled chondrule precursors have a large viscosity, the critical velocity for collisional sticking/merging could be as large as $1\ {\rm km}\ {\rm s}^{-1}$ when the temperature of supercooled droplets is below $1400$--$1500\ {\rm K}$ (Figure \ref{fig2}).
  \item Behind the shock front of the shock wave, recondensation of evaporated fine dust grains occurs before the temperature of supercooled precursors drops below the glass transition temperature, and these supercooled precursors can avoid turning into glassy chondrules (Figure \ref{fig4}).
  \item The expected number of collisions for submillimeter-sized chondrules is lower than unity when we assume $\rho_{{\rm c}, 0} / \rho_{{\rm g}, 0} = 2 \times 10^{-3}$; therefore, most of the chondrule precursors that are heated above their liquidus temperature turn into supercooled droplets and can maintain their supercooling state until the recondensation of fine dust grains occurs.
Conversely, millimeter-sized large chondrules collide frequently, and for the case of large-scale shock waves with $L \gg l_{\rm stop}$, most of the millimeter-sized chondrules have experienced collision when $\rho_{{\rm c}, 0} / \rho_{{\rm g}, 0} \gtrsim 2 \times 10^{-3}$ (Figure \ref{fig7}).
  \item With respect to the survivability of crystallized chondrules, shock waves with a spatial scale of $L \gtrsim 10^{4}\ {\rm km}$ may be desirable because the impact velocity of chondrules is inversely proportional to the spatial scale of the shock wave (Section \ref{sec3.5}).
\end{enumerate}

\acknowledgments

We thank the anonymous referee for thoughtful comments.
This work is supported by JSPS KAKENHI Grant (JP15K05266; JP18K03721).
S.A.\ is supported by the Grant-in-Aid for JSPS Research Fellow (JP17J06861).

\appendix

\section{Droplet--droplet collision experiments}
\label{ddc}

\begin{figure}
\centering
\includegraphics[width=0.5\columnwidth]{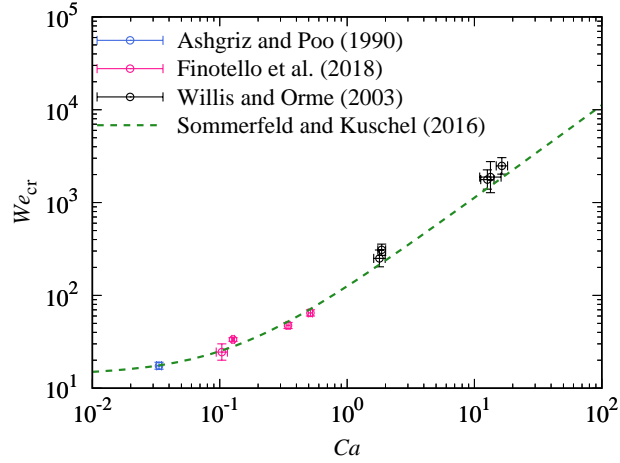}
\caption{
Experimental data of head-on collisions for equal-sized droplets \citep{Ashgriz+1990,Willis+2003,Finotello+2018a} and the proposed equation of the critical Weber number \citep{Sommerfeld+2016}.
In these experiments, the Ohnesorge number ${\rm Oh}$ is given and the range of the critical Weber number ${\rm We}_{\rm cr}$ is reported.
Then, we can evaluate the capillary number ${\rm Ca}$ from ${\rm Ca} = {\rm Oh} \sqrt{{\rm We}_{\rm cr}}$.
}
\label{fig11}
\end{figure}

The dynamics of droplet--droplet collisions has been studied for a long time because of its complexity as a fluid dynamics phenomenon.
In particular, understanding the effect of viscosity and surface energy on binary droplet collisions is of great importance for understanding the outcomes of binary equal-sized droplet collision.
The dynamics of binary equal-sized droplet collision has been investigated by numerous experimental and numerical studies \citep[e.g.,][]{Ashgriz+1990,Finotello+2017}.
\citet{Sommerfeld+2016} proposed the criteria for collisional sticking as follows:
\begin{equation}
{\rm We}_{\rm cr} = 111.66 {\rm Ca} + 13.89.
\label{eqSK2016b}
\end{equation}
In Figure \ref{fig11}, we checked the validity of the formula given by \citet{Sommerfeld+2016} by using the experimental data reported by \citet{Ashgriz+1990}, \citet{Willis+2003}, and \citet{Finotello+2018a}.

Several previous studies \citep[e.g.,][]{Qian+1997,Gotaas+2007} have proposed utilizing the dependence of ${\rm We}_{\rm cr}$ on the Ohnesorge number ${\rm Oh}$.
The Ohnesorge number ${\rm Oh}$ is given by
\begin{equation}
{\rm Oh} \equiv \frac{\eta}{\sqrt{2 \rho \sigma r}} \equiv \frac{{\rm Ca}}{\sqrt{\rm We}}.
\end{equation}
\citet{Gotaas+2007} proposed a relationship between ${\rm We}_{\rm cr}$ and ${\rm Oh}$ as follows:
\begin{equation}
{\rm We}_{\rm cr} = \left\{ \begin{array}{ll}
            14.8 + 643.1 {\rm Oh} & {({\rm Oh} < 0.04)}, \\
            9309 {\rm Oh}^{1.7056} & {({\rm Oh} \geq 0.04)}.
            \end{array} \right.
\label{eqG+2007}
\end{equation}
We can rewrite the latter part of Equation (\ref{eqG+2007}) by using ${\rm Ca}$ instead of ${\rm Oh}$:
\begin{equation}
{\rm We}_{\rm cr} = 138.7 {\rm Ca}^{0.9206}.
\label{eqG+2007c}
\end{equation}
The coefficient and the exponent in Equation (\ref{eqG+2007c}) are quite close to the coefficient and the exponent in the first term of Equation (\ref{eqSK2016b}).
In addition, both Equations (\ref{eqSK2016b}) and (\ref{eqG+2007}) asymptote to ${\rm We}_{\rm cr} \simeq 14$ for the inviscid limit (${\rm Ca} \to 0$ and ${\rm Oh} \to 0$).
These facts support the validity of the criteria for collisional sticking proposed by \citet{Sommerfeld+2016}.

Recently, \citet{Li+2016} investigated the collisions of two droplets with different viscosities, and they revealed that penetration and encapsulation are the typical outcomes for droplet collisions with a high relative viscosity ratio.
These collision outcomes may have the potential to form enveloping compound chondrules.
Our numerical results also suggest that collision of chondrule precursors frequently occurs with two precursors with different temperature, i.e., different viscosities (see Figure \ref{fig4}).

Compared with equal-size droplet collisions, unequal-size droplet collisions are more relevant to the practical situation of compound chondrule formation.
For the case of collision with low-viscosity droplets, \citet{Ashgriz+1990} and \citet{Tang+2012} found that the critical impact velocity significantly increases as the size ratio $\Delta \equiv r_{\rm small} / r_{\rm large}$ decreases, where $r_{\rm small}$ and $r_{\rm large}$ are the radii of smaller and larger droplets, respectively.
This size-ratio dependence of the critical Weber number may be due to the decrease in the relative kinetic energy determined by the total mass, and a theoretical model that is based on energy balance generally reproduces the experimental trend \citep{Tang+2012}.
Although we expect that this trend is also shown for collisions between highly viscous droplets, we have no reliable experimental data yet.
Future studies on this topic are therefore essential.

\section{Droplet--solid collision experiments}
\label{dsc}

The outcome of a droplet impact on a solid surface also depends on the physical properties of the liquid, and there have been several studies on the sticking/splashing criteria of a droplet--solid collision \citep[e.g.,][]{Walzel1980,Mundo+1995,Josserand+2016}.
Considering the equations of energy conservation, \citet{Mundo+1995} analytically derived the criteria for collisional sticking/splashing as follows \citep[see also][]{Chandra+1991}:
\begin{equation}
{\rm We}_{\rm cr} = \frac{9}{2} \beta^{4} {\rm Ca} + 3 {\left( 1 - \cos{\Theta} \right)} \beta^{2} - 12,
\label{eqM95}
\end{equation}
where $\beta$ is the maximum spreading diameter of the droplet scaled with the initial diameter and $\Theta$ is the contact angle.
\citet{Chandra+1991} revealed that the maximum spreading diameter is $\beta \simeq 2$--$3$, and this relation matches the experimentally obtained correlation between ${\rm We}_{\rm cr}$ and ${\rm Ca}$ \citep{Mundo+1995}.
This equation is a special case of Equation (\ref{eqDA}), implying that the energy dissipation mechanism in droplet--solid collisions may be similar to that of a droplet--droplet collision.
In addition, the equivalent critical Reynolds number is ${\rm Re}_{\rm cr, v} = {(9 / 2)} \beta^{4} \sim 10^{2}$, which is similar to the critical Reynolds number for droplet--droplet collisions.

We acknowledge, however, that the physics of droplet--solid collisions is still not well-understood.
Therefore, we roughly evaluate the sticking/splashing criteria of droplet--solid collision by using Equation (\ref{eqVcr}) instead of Equation (\ref{eqM95}), and our estimate is no more than an order estimation.
We will study droplet--solid collisions by using hydrodynamics simulations in the future.

\bibliographystyle{aasjournal}

\bibliography{Compound}





\end{document}